\documentclass[journal]{IEEEtran}
\IEEEoverridecommandlockouts
\usepackage{amsmath,amsfonts}
\usepackage{bm}
\usepackage{amssymb}
\usepackage{hhline}
\usepackage{algorithmic}
\usepackage{algorithm}
\usepackage{array}
\usepackage[dvipsnames]{xcolor}
\usepackage{hyperref}
\ifCLASSOPTIONcompsoc
    \usepackage[caption=false, font=normalsize, labelfont=sf, textfont=sf]{subfig}
\else
\usepackage[caption=false, font=footnotesize]{subfig}
\fi
\usepackage{textcomp}
\usepackage{stfloats}
\usepackage{url}
\usepackage{verbatim}
\usepackage{graphicx}
\usepackage{cite}
\newcommand{\norm}[1]{\left\lVert#1\right\rVert}

\newcolumntype{P}[1]{>{\centering\arraybackslash}p{#1}}
\mathchardef\mhyphen="2D

\begin{document}

\title{Cut-FUNQUE: An Objective Quality Model for Compressed Tone-Mapped High Dynamic Range Videos}

\author{Abhinau K. Venkataramanan, Cosmin Stejerean, Ioannis Katsavounidis,\\ Hassene Tmar, and Alan C. Bovik, ~\IEEEmembership{Life Fellow,~IEEE}
\thanks{This research was sponsored by a grant from Meta Video Infrastructure, and by grant number 2019844 from the National Science Foundation AI Institute for Foundations of Machine Learning (IFML).}}

\markboth{Journal of \LaTeX\ Class Files,~Vol.~XX, No.~X, XXXX~20XX}%
{Venkataramanan \MakeLowercase{\textit{et al.}}: One Transform to Compute Them All}

\maketitle

\begin{abstract}
High Dynamic Range (HDR) videos have enjoyed a surge in popularity in recent years due to their ability to represent a wider range of contrast and color than Standard Dynamic Range (SDR) videos. Although HDR video capture has seen increasing popularity because of recent flagship mobile phones such as Apple iPhones, Google Pixels, and Samsung Galaxy phones, a broad swath of consumers still utilize legacy SDR displays that are unable to display HDR videos. As result, HDR videos must be processed, i.e., tone-mapped, before streaming to a large section of SDR-capable video consumers. However, server-side tone-mapping involves automating decisions regarding the choices of tone-mapping operators (TMOs) and their parameters to yield high-fidelity outputs. Moreover, these choices must be balanced against the effects of lossy compression, which is ubiquitous in streaming scenarios. In this work, we develop a novel, efficient model of objective video quality named Cut-FUNQUE that is able to accurately predict the visual quality of tone-mapped and compressed HDR videos. Finally, we evaluate Cut-FUNQUE on a large-scale crowdsourced database of such videos and show that it achieves state-of-the-art accuracy. 
\end{abstract}


\section{Introduction}
\label{sec:introduction}
The Human Visual System (HVS) encounters a diverse range of luminances, or brightness, in real-world scenarios, spanning from the faint glow of starlight at 0.0003~cd/\(\text{m}^2\) (nits) to the intense brightness of sunlight reaching up to 30,000 nits on a clear day. Thanks to the adjustment of the pupil size by the iris to varying brightness levels, the HVS can perceive an extensive range of luminances, ranging from approximately \(10^{-6}\) nits to \(10^8\) nits. However, conventional imaging and display systems are limited to capturing or producing narrow ranges of luminances, typically up to about 100 nits. These systems, categorized as low or standard dynamic range (SDR) systems, are also only capable of capturing or displaying about 35\% of the visible color gamut. Notable examples of legacy SDR standards include sRGB \cite{ref:srgb} and ITU BT. 709 \cite{ref:rec_709}.

Over the years, the development of high dynamic range (HDR) imaging has aimed to better align imaging and display systems with the capabilities of the HVS. Contemporary HDR standards, exemplified by ITU BT. 2100 \cite{ref:rec_2100}, have the capacity to capture luminances spanning from \(10^{-4}\) to \(10^4\) nits, along with a wide color gamut (WCG) that encompasses approximately 75\% of the visible color volume. To achieve this, HDR imaging employs computational imaging techniques that blend two or more images taken at different exposure settings. Novel ``optoelectrical transfer functions'' (OETFs) extend the conventional notion of ``gamma'' from legacy Cathode Ray Tube (CRT) displays to effectively encode and transmit this wide range of brightnesses.

In particular, the BT. 2100 standard incorporates two encoding functions, namely the Perceptual Quantizer (PQ) \cite{ref:pq} and the Hybrid Log-Gamma (HLG) \cite{ref:hlg}. PQ is engineered as a ``forward-compatible'' standard, capable of encoding luminances up to \(10^{4}\) nits, and is commonly utilized by professional studios to deliver high-quality HDR content. Notably, the PQ encoding function serves as the foundation of standards like HDR10 \cite{ref:hdr10} and HDR10+ \cite{ref:hdr10_plus}. In contrast, HLG is designed to be ``backward-compatible'' with SDR standards by incorporating a ``gamma" curve similar to that used in SDR, in the SDR luminance range. While the HLG standard does not explicitly define a peak luminance, a nominal value of 1000 nits is often adopted. Due to its backward compatibility, HLG has gained adoption by satellite TV networks for the delivery of HDR content \cite{ref:hlg_satellite}. Notably, the emerging Dolby Vision standard \cite{ref:dolbyvision} supports both PQ and HLG encoding functions.

The widespread streaming of HDR videos faces a significant challenge due to the limited availability of true HDR displays. According to the BT. 2100 standard, true HDR systems are defined as those capable of supporting a minimum of 1000 nits \cite{ref:rec_2100}. However, a majority of budget-friendly HDR displays fall short of this criterion, reaching a peak luminance of 800 nits or less \cite{ref:hdr_tv_survey}. In fact, a substantial portion of existing displays are only capable of supporting SDR formats. Consequently, to ensure accessibility of HDR video content to a broader consumer base, it becomes essential to ``down-convert'' them to the SDR range, a process commonly referred to as ``tone-mapping.''

Section \ref{sec:evaluation_live_tmhdr} provides a brief overview of various tone-mapping methods proposed in the literature. However, the inherent limitations of SDR systems lead to distortions during the tone-mapping process. These distortions typically manifest as either contrast reduction or enhancement, as well as the potential losses or amplification of details in dark or bright regions. Furthermore, the remapping of color across dynamic ranges and the wide color gamut (WCG) employed by high dynamic range (HDR) can give rise to chromatic distortions, including hue shifts and chroma-clipping artifacts \cite{ref:cadik} \cite{ref:gma_review}. Additionally, the necessity for lossy compression when streaming videos over the internet introduces another layer of distortions, including flattening of details, blocking in regions containing high motion or texture and banding on smoother regions.

Here, we tackle the problem of objective quality assessment of compressed tone-mapped videos. In particular, we design the new Cut-FUNQUE objective quality model which is based on the recently developed FUNQUE \cite{ref:funque} \cite{ref:funque_plus} framework, and demonstrate its effectiveness in automatically predicting the subjective quality of tone-mapped videos. Cut-FUNQUE is the sum total of three key novel contributions.
\begin{enumerate}
    \item The first contribution is the development of a novel perceptually uniform encoding of color signals (\textbf{PUColor}) that we use to represent both HDR and SDR color stimuli in a common domain. In this manner, PUColor enables the meaningful comparison of stimuli across dynamic ranges, which is essential when comparing HDR and SDR videos.
    \item Secondly, Cut-FUNQUE utilizes a \textbf{binned-weighting} approach to separately handle image regions having different visual characteristics such as brightness, contrast, and temporal complexity.
    \item Finally, Cut-FUNQUE also utilizes novel \textbf{statistical similarity} measures of visual quality to overcome the limitations of pixel-wise comparisons across dynamic ranges.
\end{enumerate}

We demonstrate the efficacy of Cut-FUNQUE by conducting evaluations on the recently introduced LIVE Tone-Mapped HDR (LIVE-TMHDR) subjective database.

The remainder of this manuscript is organized as follows. Section \ref{sec:background} provides an overview of the literature regarding the objective quality assessment of tone-mapped images and videos. Section \ref{sec:cut_funque} describes the proposed Cut-FUNQUE model in detail, and in Section \ref{sec:evaluation}, we report the results of comparing Cut-FUNQUE against existing video quality models in the literature.

Finally, we present a summary of our findings and directions for future work in Section \ref{sec:conclusion}.
\section{Background}
\label{sec:background}

Due to the longstanding nature of the HDR tone-mapping problem, several algorithms have been developed to assess the quality of tone-mapped media, particularly images. The Dynamic Range Independent Quality Assessment (DRIQA) model \cite{ref:driqa} was the first attempt at comparing HDR and SDR images despite their different dynamic ranges. The model generates distortion maps using contrast sensitivity models, cortex transforms, and psychometric probability models to measure loss, gain, and reversal of visible contrast. However, by design, the algorithm does not provide accurate ``single number'' predictions of visual quality from the distortions maps (such as by averaging) \cite{ref:driqa}.

The Tone Mapped Quality Index (TMQI) \cite{ref:tmqi} was the first quality model designed to conduct tone-mapped image quality prediction. It introduced the popular framework of hybrid Full Reference (FR) - No Reference (NR) methods. TMQI combines measures of structural fidelity, similar to Structural Similarity (SSIM) \cite{ref:ssim}, and naturalness, which may be considered an NR model, to predict overall quality. A variant of TMQI, called TMQI-NSS \cite{ref:attention_tmqi}, incorporates visual attention into the structure term and uses Natural Scene Statistics (NSS) to quantify naturalness. The NSS model borrows from NR algorithms such as BRISQUE \cite{ref:brisque} and NIQE \cite{ref:niqe}.

The FSITM \cite{ref:fsitm} algorithm is a modified version of the FSIM \cite{ref:fsim} quality model, and it is based on analyzing the local phase of complex log-Gabor wavelet transform coefficients. Binarized phase values from the reference HDR and test SDR images are compared to obtain a structural similarity measure, while naturalness is defined as the same similarity metric applied between the SDR image and the log of the HDR image. Combining this model with TMQI has been shown to improve performance \cite{ref:fsitm}. FFTMI \cite{ref:fftmi} is a fusion-based model built on this principle, and its ``atom quality models'' include FSITM computed from RGB channels, TMQI, and a no-reference model of naturalness \cite{ref:fftmi_naturalness}. These atom features are combined to predict a single quality score using a regressor.

Finally, the Tone Mapped Video Quality Index (TMVQI) \cite{ref:tmvqi} is an adaptation of the TMQI model. TMVQI uses an updated contrast sensitivity model to modulate contrast estimates, and an updated naturalness model that was estimated from HDR video frames.

Among no-reference models, the Blind TMQI (BTMQI) \cite{ref:btmqi} model utilizes entropy estimates from simulated multi-exposure images to predict structural fidelity and borrows its naturalness measure from TMQI. HIGRADE \cite{ref:higrade} employs a Natural Scene Statistics (NSS) approach similar to algorithms such as BRISQUE \cite{ref:brisque} by modeling the statistics of Mean Subtracted Contrast Normalized coefficients. In addition, HIGRADE incorporates structural information by modeling the statistics of gradient structure tensors.

The RcNet \cite{ref:rcnet} model is a deep-learning NR quality model that uses a suite of Convolutional Neural Networks to predict the DRIQA distortions maps without using a reference HDR source. The distributions of MSCN coefficients of the SDR image and the predicted distortion maps are used to predict its quality. The state-of-the-art deep method is the Multi-Scale Multi-Layer \cite{ref:msml} model, which uses pooled features extracted from the intermediate layers of a pretrained ResNet-50 network. The outputs of three hidden layers are concatenated to yield a feature vector containing 9216 features, and partial least squares (PLS) is used to reduce the final dimensionality of the feature vector to fifteen.

\section{Cut-FUNQUE}
\label{sec:cut_funque}

\begin{figure}[t]
    \centering
    \includegraphics[width=\linewidth]{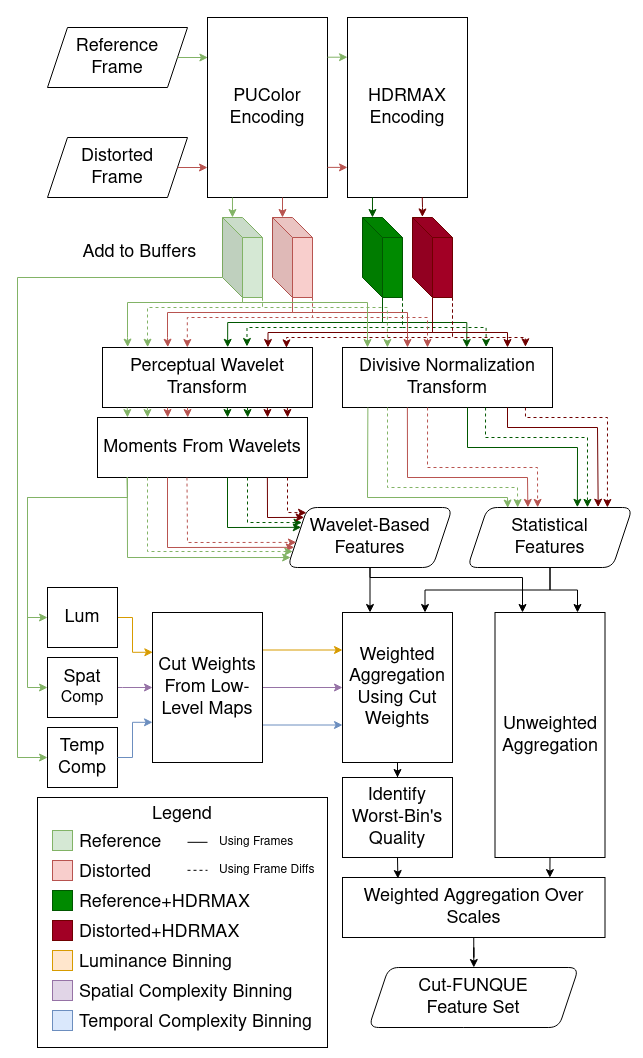}
    \caption{An overview of the Cut-FUNQUE quality prediction model.}
    \label{fig:cut_funque_flowchart}
\end{figure}

As described in Section \ref{sec:background}, successful models of tone-mapped video quality are hybrid ones that include both full-reference and no-reference features, such as the structure and naturalness terms in TMQI \cite{ref:tmqi}. Moreover, with the increase in available dataset sizes, feature-based models have emerged as a powerful tool for quality modeling, both for no-reference models such as V-BLIINDS \cite{ref:v_bliinds}, VIDEVAL \cite{ref:ugc_vqa}, and HIGRADE \cite{ref:higrade}, and full-reference methods such as VMAF \cite{ref:vmaf} and the FUNQUE framework \cite{ref:funque_plus} \cite{ref:hdr_funque_plus}. Therefore, we employ a fusion-based bag-of-features approach consisting of smaller ``atom'' quality models and features designed based on perceptual principles.

To design atom features appropriate for the quality assessment of tone-mapped videos, we first made a set of observations regarding the challenges faced by prior work in the area, and the effects of tone-mapping and compression. The primary challenge that is endemic to the field of tone-mapped video quality assessment is identifying a common domain in which both HDR and SDR may be well represented. Based on prior work in the area of perceptually uniform representations \cite{ref:pu21}, we posited that a good transform domain designed for this purpose would be \textbf{perceptually uniform over a wide range of brightnesses}.

While PU21 \cite{ref:pu21} is a promising choice for such a transform, it was derived based on a luminance contrast sensitivity model. Since color distortions are an important aspect of tone mapping, an ideal transform would be a perceptually uniform color encoding function (PUColor) that can represent three-channel color values. We describe our proposed PUColor encoding function in Section \ref{sec:cut_funque_pucolor}.

Furthermore, we make the following observations regarding the nature of distortions that arise from tone-mapping and compression.
\begin{itemize}
    \item Since tone-mapping involves applying compressive or sigmoidal non-linearities, \textbf{contrast changes at various luminance levels may be different}. These non-linearities are typically responsible for distortions, such as loss of detail/contrast, in very bright and dark regions of a video. On the other hand, while the contrast in mid regions is typically preserved, distortions in the mid regions may be indicative of global brightness changes.
    \item \textbf{Loss/gain of detail in various regions may be affected by both tone-mapping and compression.} Compression alters local contrast by inducing blockiness (which is blur-like) in textured (high-contrast) regions and banding in smooth (low-contrast) regions. On the other hand, local processing in tone-mapping often boosts regions of low local contrast and attenuates regions of high local contrast. Therefore, the nature of distortions also varies with the amount of local contrast.
    \item \textbf{Regions with high temporal variation are also affected by both tone-mapping and compression}, in the form of flickering and misalignment/blocking due to the quantization of motion vectors.
\end{itemize}

To incorporate this prior knowledge into the algorithm, we partitioned each video frame into patches (referred to as ``cuts'' since they are non-overlapping) and (softly) classified each cut into various ``bins'' depending on their low-level spatio-temporal properties. We then employed a weighted aggregation method to pool quailty features within each bin to characterize different ``types'' of frame regions. Finally, we identified the (predicted) worst-quality regions to account for quality-based saliency. This procedure is described in further detail in Section \ref{sec:cut_funque_weighted_bins}. Furthermore, we combined the proposed PUColor with the recently developed HDRMAX transform \cite{ref:hdrmax}, which emphasizes quality prediction on bright and dark regions of input frames, which are perceptually important for HDR stimuli. 

To enable effective quality prediction, we extracted various quality-aware features from input videos. These features included spatial and temporal models of quality such as SSIM \cite{ref:ssim}, VIF \cite{ref:funque_plus}, DLM \cite{ref:dlm}, and ST-RRED \cite{ref:strred}. Notably, all the aforementioned quality models are ``pixel-based'' methods that directly compare (processed) pixel values. Such methods are limited in their capacity to accurately compare video frames across dynamic ranges. 

For example, the SSIM score between a pair of image regions is unity only if their pixel values are identical. However, since pixel values of an SDR frame occupy a limited range, they are likely to be significantly different in value from their HDR frame counterparts, irrespective of the quality of tone-mapping. To overcome this, we propose novel statistical similarity (StSIM) measures of quality, based on principles of Natural Scene Statistics (NSS). StSIM features compare the distributions of normalized pixel values, rather than comparing pixel values directly. This makes StSIM features more robust to changes in dynamic range. 

The feature set employed by Cut-FUNQUE is described in further detail in Sections \ref{sec:cut_funque_wavelet_feature_extraction} and \ref{sec:cut_funque_nss_feature_extraction}. A flowchart illustrating the Cut-FUNQUE processing flow is presented in Figure \ref{fig:cut_funque_flowchart}.

\subsection{A Perceptually Uniform Encoding Function for Color}
\label{sec:cut_funque_pucolor}
In this section, we present the derivation of a perceptually uniform color encoding function (PUColor) using a ``post-receptoral'' model of the spatio-chromatic contrast sensitivity function (SCCSF) \cite{ref:omni_csf}. This SCCSF model has been used in prior work to derive perceptually uniform luminance encoding functions \cite{ref:pu21}. Here, we design the PUColor encoding function to be approximately uniform along three ``chromatic'' directions, which we choose to correspond to opponent color channels - achromatic, red-green, and blue-yellow. 

If \(t\) denotes the detection threshold, i.e., a Just-Noticeable-Difference (JND), we say that the \(PUColor\) encoding function of LMS values \(\mathbf{x}~=~[x_L, x_M, x_S]^T\) is uniform along three ``chromatic'' vectors \(\mathbf{u_i}\) if
\begin{equation}
    PUColor(\mathbf{x} + t(\mathbf{x}, \mathbf{u_i}) \cdot \mathbf{u_i}) - PUColor(\mathbf{x}) \approx C \mathbf{e_i},
    \label{eq:pucolor_def}
\end{equation}
where \(C\) is an arbitrary constant and the set \(\{\mathbf{e_i}\}\) refers to the standard basis vectors. Here, ``LMS'' refers to a color space that represents the responses of cone cells in the retina in the long, medium, and short-wavelength regions. In the analysis below, we use \(C=1\) for convenience. We then scale the derived function such that the range of encoded values is in the range \([0, 1]\).

Using a first-order approximation of \(PUColor\), the uniformity condition in Eq. \ref{eq:pucolor_def} may be rewritten as
\begin{equation}
    \mathbf{J}(\mathbf{x}) \mathbf{u_i} \approx \frac{1}{t(\mathbf{x}, \mathbf{u_i})}\mathbf{e_i},
    \label{eq:jacobian_eig}
\end{equation}
where \(\mathbf{J}(\mathbf{x})\) denotes the Jacobian matrix of the partial derivatives of \(PUColor\). It may be observed that Eq. \ref{eq:jacobian_eig} is a generalized eigenvalue relationship between the threshold function and the direction of uniformity. Collecting the chromatic vectors into a matrix \(\mathbf{U} = \left[\mathbf{u_1}, \mathbf{u_2}, \mathbf{u_3}\right]\), then

\begin{equation}
    \mathbf{J}(\mathbf{x}) =  \textit{diag}\left(t^{-1}\left(\mathbf{x}, \mathbf{u}_i\right)\right)\mathbf{U}^{-1},
\end{equation}
where \(\textit{diag}\) denotes a diagonal matrix.

To complete the description of the Jacobian matrix, the threshold function \(t\) must be characterized. Here, we utilized the threshold function proposed in the SCCSF model \cite{ref:omni_csf}, which was defined in terms of a transformation matrix \(\mathbf{M_{ARB}}\), three ``base sensitivity functions'' \(s_A(\cdot), s_R(\cdot), s_B(\cdot)\), and spatial frequency \(\rho\)

\begin{equation}
    t(\mathbf{x}, \mathbf{u}) = \min_\rho \norm{\left(\frac{\mathbf{s_{ARB}}(\rho, x_L + x_M)}{x_L + x_M}\right) \odot \left(\mathbf{M_{ARB}} \mathbf{u}\right)}_2^{-1},
    \label{eq:thresh}
\end{equation}
where \(\odot\) denotes the elementwise (Hadamard) product.

Finally, to obtain the \(PUColor\) encoding function, we enforced the boundary condition that \(PUColor(\mathbf{0}) = 0\) and integrated the Jacobian over the straight line \(z = \{ \lambda\mathbf{x}; \lambda \in [0, 1]\}\). Thence,
\begin{align}
    PUColor(\mathbf{x}) &= \int\limits_0^1 \mathbf{J}\left(\mathbf{z}\left(\lambda\right)\right) \mathbf{z}'\left(\lambda\right) d\lambda \\
    &= \left(\int\limits_0^1 \frac{d\lambda}{t\left(\lambda\mathbf{x}, \mathbf{u}_i\right)} \right) \odot \left(\mathbf{U}^{-1}\mathbf{x}\right).
    \label{eq:pucolor_ints}
\end{align}

The three integrals, corresponding to the three chromatic directions \(\mathbf{u_i}\), were evaluated numerically. A closer look at Eq. \ref{eq:thresh} reveals that \(t\) depends on \(\mathbf{x}\) only through \(y = x_L + x_M\), which denotes the luminance. Consequently, the integrals in Equation \ref{eq:pucolor_ints} share the same property. Therefore, to apply the PUColor encoding in practice, we approximate the results of the numerical integrals using non-linear functions of the form \(h(y) / y\), where

\begin{equation}
    h(y; p) = \left(\frac{p_1 + p_2 y^{p_4}}{1 + p_3 y^{p_4}}\right)^{p_5}.
    \label{eq:five_param_nonlinearity}  
\end{equation}

\begin{figure}[t]
    \centering
    \includegraphics[width=0.8\linewidth]{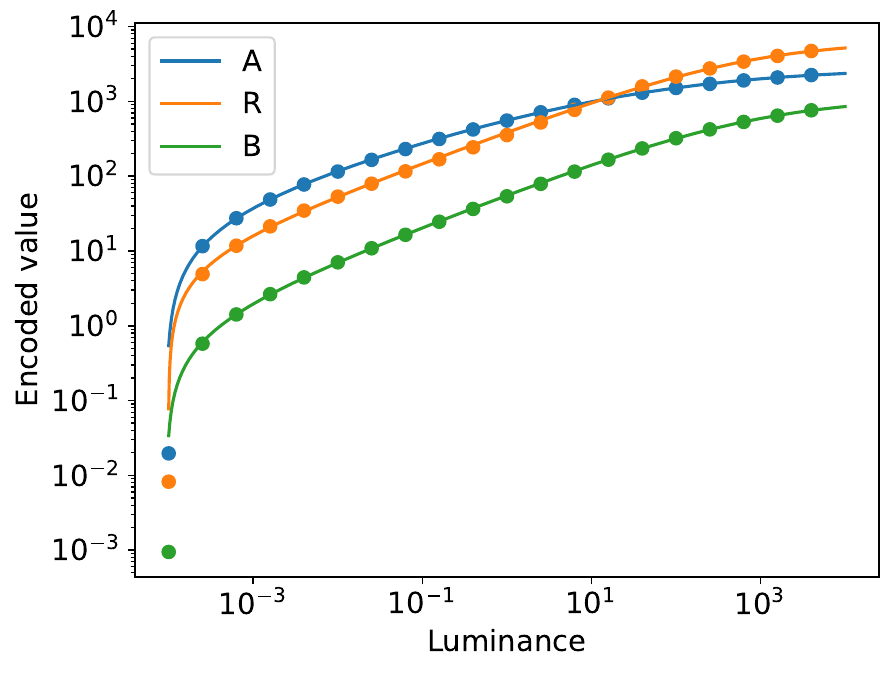}
    \caption{Numerical integrals of the threshold function and approximations using non-linear functions}
    \label{fig:pucolor_int_num_fits}
\end{figure}

A comparison of the values of numerical integrals and the corresponding non-linear function fits for various luminance values is presented in Figure \ref{fig:pucolor_int_num_fits}. All three numerical fits achieved \(r^2\) values over 0.999. To ensure that the three ``chromatic directions'' correspond to an opponent representation of color, we chose \(\mathbf{U} = \mathbf{M_{DKL}}^{-1}\), where \(\mathbf{M_{DKL}}\) is the transformation matrix corresponding to the Derrington-Krauskopf-Lennie (DKL) opponent color space \cite{ref:dkl}. Therefore, the final PUColor encoding function has the simple form
\begin{equation}
    PUColor(\mathbf{x}) = \frac{\mathbf{h_{ARB}}(x_L + x_M)}{x_L + x_M} \odot \left(\mathbf{M_{DKL}} \cdot \mathbf{x}\right).
\end{equation}

The encoded values consist of one achromatic channel (\(L\)) and two chromatic channels that may be denoted by \(a\) and \(b\), analogous to color spaces such as CIELAB \cite{ref:cielab}. The two chromatic channels encode color together, where the ``angle'' between the two components \(\arctan\left(b/a\right)\) encodes the hue, and the magnitude \(\sqrt{a^2 + b^2}\) encodes chromaticity. Therefore, we combine the two channels into one complex-valued color channel \(c = a + jb\), where \(j = \sqrt{-1}\), and the argument (\(\arg{c}\)) and magnitude (\(|c|\)) of \(c\) represent the hue and chromaticity respectively.
\subsection{HDRMAX Encoding}
\label{sec:cut_funque_hdrmax}
HDRMAX \cite{ref:hdrmax} refers to a non-linear pre-processing method that seeks to emphasize perceptually important image regions in HDR videos. In particular, HDRMAX emphasizes the measurement of distortions in bright and dark image regions. This is achieved by pre-processing image frames in two steps - local normalization and a point non-linear transform.

Local min-max normalization is used to first normalize local luma (and chroma) values within overlapping \(17\times17\) neighborhoods to the range \([-1, 1]\):
\begin{equation}
    I_{minmax}(i, j) = 2\frac{I(i,j) - I_{min}(i, j)}{I_{max}(i,j) - I_{min}(i, j)}.
\end{equation}

Normalized luma (and chroma) values are then subjected to a double-exponential non-linearity
\begin{equation}
    \mathrm{HDRMAX}(x) = \mathrm{sgn}(x)\left(e^{4|x|}-1\right).
\end{equation}

Although HDRMAX was designed to enhance the prediction of the quality of HDR stimuli \cite{ref:hdr_funque_plus}, it has also been effective for SDR video quality prediction \cite{ref:hdrmax}. So, we adopt it as a pre-processing step for both HDR reference and SDR (i.e., tone-mapped) test videos.
\subsection{Perceptually-Sensitized Wavelet Transform}
\label{sec:cut_funque_wavelet_transform}
Following the FUNQUE framework \cite{ref:funque} \cite{ref:funque_plus}, we compute a perceptually-sensitized wavelet decomposition of all analyzed reference and test frames after applying the PUColor and HDRMAX transformations. The Self Adaptive Scale Transform (SAST) \cite{ref:sast} was first applied to account for viewing distance, which was assumed to be three times the display height as in \cite{ref:funque_plus}. This assumption corresponds to typical HD TV viewing distances.

Then, a Haar wavelet transform was applied to obtain a band-pass decomposition of the input frames. To account for the visibility of various artifacts due to the contrast sensitivity of the visual system, the subbands of the Haar decomposition were weighted using Watson's model of wavelet subband contrast sensitivity \cite{ref:watson}. We preferred Watson's model of contrast sensitivity over other models described in \cite{ref:funque_plus} due to its wavelet domain definition and its modeling of chroma visibility.

The perceptually-sensitized wavelet decomposition was used in \cite{ref:funque_plus} to compute multi-scale local moments (i.e., mean, variance, and covariance) of the reference and test frames, which were used to compute Multi-Scale Enhanced SSIM (MS-ESSIM) \cite{ref:essim}. We adopt a similar procedure here, but we reuse the same moments to compute three sets of quality features - ESSIM, Visual Information Fidelity (VIF) \cite{ref:vif}, and Spatio-Temporal Reduced Reference Entropic Differences (ST-RRED) \cite{ref:strred}. Moreover, local moments of the reference frame were also used to compute cut-assignment weights, which are described in Section \ref{sec:cut_funque_weighted_bins}. Such extensive computation sharing follows the FUNQUE framework and contributes to Cut-FUNQUE's efficiency despite the large number of features.

Let \(x\) and \(y\) denote the reference and test images, and \(X_{l, \lambda, \theta}\) and \(Y_{l, \lambda, \theta}\) denote the wavelet decompositions of their luma channels, where \(\lambda \in \{1 \dots L\}\) denotes the wavelet level (i.e., ``scale'') and \(\theta \in \{A, H, V, D\}\) denote the ``orientation'' of the subband. Then, local moments of the luma channel within non-overlapping windows of size \(2^\lambda\times2^\lambda\) are computed as
\begin{equation}
    \mu_{l, x, \lambda}(i, j) = 2^{-\lambda}X_{l, \lambda, A}(i, j),
\end{equation}
\begin{equation}
    \mu_{l, y, \lambda}(i, j) = 2^{-\lambda}Y_{l, \lambda, A}(i, j),
\end{equation}
\begin{equation}
    \sigma^2_{l, x, \lambda}(i, j) = 2^{-2\lambda}\sum_{k=1}^{\lambda} \sum_{P^k_{ij}} \sum_{\{H, V, D\}} X_{l, k, \theta}^2(m, n),
    \label{eq:sig_x}
\end{equation}
\begin{equation}
    \sigma^2_{l, y, \lambda}(i, j) = 2^{-2\lambda}\sum_{k=1}^{\lambda} \sum_{P^k_{ij}} \sum_{\{H, V, D\}} Y_{l, k, \theta}^2(m, n),
\end{equation}
\begin{equation}
    \sigma_{l, xy, \lambda}(i, j) = 2^{-2\lambda}\sum_{k=1}^{\lambda} \sum_{P^k_{ij}} \sum_{\{H, V, D\}} X_{l, k, \theta}(m, n) Y_{l, k, \theta}(m, n),
\end{equation}
where \(P^k_{ij} = \{(m, n) \mid i2^{\lambda-k} \leq m < (i+1)2^{\lambda-k}, j2^{\lambda-k} \leq n < (j+1)2^{\lambda-k}\}\) denotes disjoint \(2^{\lambda-k}\times2^{\lambda-k}\) blocks.

Analyzing Eq. \eqref{eq:sig_x}, it may be seen that multi-scale variances can be computed iteratively using the relationship

\begin{align}
    \sigma_{l, x, \lambda+1}^2(i,j) = 2^{-2}\sum\limits_{m=i}^{i+1}\sum\limits_{n=j}^{j+1}\sigma_{l, x, \lambda}(m,n)^2 + \Tilde{\sigma}^2_{l, x, \lambda+1}(i,j),
    \label{eq:iterative_var}
\end{align}
where
\begin{align}
    \Tilde{\sigma}^2_{l, x, \lambda+1}(i,j) = 2^{-2(\lambda+1)} \sum_{\{H, V, D\}} X^2_{l, \lambda+1, \theta}(i, j).
\end{align}
The energies \(\sigma_{l, x, \lambda+1}^2(i,j)\), and \(\sigma_{l, xy, \lambda+1}(i,j)\) were also computed iteratively in a similar manner.

Local moments of the complex-valued chroma channel were computed similarly, with the key difference being the use of complex conjugates when computing variance and covariance. As an example,
\begin{equation}
    \sigma_{c, xy, \lambda}(i, j) = 2^{-2\lambda}\sum_{k=1}^{\lambda} \sum_{P^k_{ij}} \sum_{\{H, V, D\}} X_{c, k, \theta}(m, n) Y^{*}_{c, k, \theta}(m, n),
\end{equation}
where \(^{*}\) denotes the complex conjugate. Once again, multi-scale moments were computed iteratively as in Eq. \eqref{eq:iterative_var}.

Using this method, we compute local moments within \(8\times8\), \(16\times16\), \(32\times32\), and \(64\times64\) windows, which correspond to the four scales at which Cut-FUNQUE is applied.
\subsection{Wavelet-Domain Feature Extraction}
\label{sec:cut_funque_wavelet_feature_extraction}
In this section, we describe the wavelet-domain features computed using the perceptually-sensitized wavelet transform described in Section \ref{sec:cut_funque_wavelet_transform}. In particular, this section describes the procedure to calculate local quality-aware features that were spatially aggregated using the method described in Section \ref{sec:cut_funque_weighted_bins}.

We computed four sets of wavelet-domain features - SSIM, VIF, RRED, and DLM, from which SSIM, VIF, and RRED were computed using the local moments described in Section \ref{sec:cut_funque_wavelet_transform}. Local SSIM scores were split into their two components, namely luminance and contrast-structure similarity \cite{ref:ssim}. Since we compute SSIM features on both luma and complex-valued chroma channels, we rename the two components as \(SSIM_{\mu}\) and \(SSIM_{\sigma}\) as general terminology we use for both channels. SSIM features from the luma channel were computed as:
\begin{equation}
    \mathrm{L\mhyphen SSIM}_{\mu}(i, j) = \frac{2\mu_{l, x}(i, j)\mu_{l, y}(i, j) + C_1}{\mu_{l, x}^2(i, j) + \mu_{l, y}^2(i, j) + C_1}
\end{equation}
\begin{equation}
    \mathrm{L\mhyphen SSIM}_{\sigma}(i, j) = \frac{2\sigma_{l, xy}(i, j) + C_2}{\sigma_{l, x}^2(i, j) + \sigma_{l, y}^2(i, j) + C_2}
\end{equation}
and the chroma SSIM features were computed as
\begin{equation}
    \mathrm{C\mhyphen SSIM}_{\mu}(i, j) = \frac{2|\mu_{c, x}(i, j)\mu_{c, y}(i, j)| + C_1}{|\mu_{c, x}^2(i, j)| + |\mu_{c, y}^2(i, j)| + C_1}
\end{equation}
\begin{equation}
    \mathrm{C\mhyphen SSIM}_{\sigma}(i, j) = \frac{2|\sigma_{c, xy}(i, j)| + C_2}{\sigma_{c, x}^2(i, j) + \sigma_{c, y}^2(i, j) + C_2}.
\end{equation}

This process was repeated across four scales, both with and without HDRMAX, yielding a total of 32 feature maps.

Visual Information Fidelity (VIF) \cite{ref:vif} is defined as the ratio of two estimates of mutual-information corresponding to the test and reference video frames. As described in \cite{ref:vif}, VIF assumes that distortions arise from a channel modeled as
\begin{equation}
    Y_{l, \lambda}(i,j) = g_{l, \lambda}(i,j) X_{l, \lambda}(i, j) + N_{l, \lambda}(i, j),
\end{equation}
where \(g_{l, \lambda}(i,j)\) is the gain of the distortion channel and \(N_{l, \lambda}(i, j) \sim \mathcal{N}(0, \sigma^2_v)\) is additive white Gaussian noise.

We adopted a similar disortion model for the complex-valued chroma channel, with the modification that all quantities may take complex values and the noise channel is a complex Gaussian, i.e., \(N_{c, \lambda}(i, j) \sim \mathcal{CN}(0, \sigma^2_v)\).

The local mutual information (MI) estimates, computed under a Gaussian assumption, were obtained from the luma channel's moments using the following equations:
\begin{equation}
    g_{l, \lambda}(i,j) = \sigma_{l, xy, \lambda}(i,j) / \sigma^2_{l, x, \lambda}(i,j),
\end{equation}
\begin{equation}
    \sigma^2_{l, v, \lambda}(i,j) = \sigma^2_{l, y, \lambda}(i,j) - g_{l, \lambda, \theta}(i,j) \sigma_{l, xy, \lambda}(i,j),
\end{equation}
\begin{equation}
    \mathrm{MI\mhyphen Test}_{l, \lambda}(i, j) = \log\left(1 + \frac{g_{l, \lambda}^2(i,j)\sigma_{l, x, \lambda}^2(i,j)}{\sigma^2_{l, v, \lambda}(i,j) + \sigma_n^2}\right),
\end{equation}
and
\begin{equation}
    \mathrm{MI\mhyphen Ref}_{l, \lambda} = \log\left(1 + \frac{\sigma_{l, x, \lambda}^2(i,j)}{\sigma_n^2}\right),
\end{equation}
where \(\sigma_n^2\) denotes the noise variance of the assumed distortion channel. Hence, local VIF scores were computed as
\begin{equation}
    \mathrm{VIF}_{l,\lambda}(i,j) = \frac{\mathrm{MI\mhyphen Test}(i, j)}{\mathrm{MI\mhyphen Ref}(i, j)}
\end{equation}

A similar procedure was used to obtain local VIF scores for the chroma channel under a complex Gaussian assumption. We omit the analogous expressions of the quantities described above for brevity, noting that differential entropy of a 1-D complex Gaussian distribution is identical to that of a 2-D real-valued Gaussian distribution having its real and imaginary parts as its components. Furthermore, we repeated the process on the differences of successive frames to obtain estimates of temporal quality, which we denote by ``TVIF''. Hence, we obtained four quality features L-VIF, C-VIF, L-TVIF, and C-TVIF, each repeated at four scales and with and without HDRMAX, yielding a total of 32 features. 

The Reduced Reference Entropic Difference (RRED) quality features \cite{ref:strred} are defined as the difference between scaled local entropy estimates, also obtained under a Gaussian assumption using the same local moments as SSIM and VIF:
\begin{equation}
   h_{l, \lambda}(i, j) = \alpha_{l, \lambda}(i,j)\log\left(2\pi e(\sigma^2_{l, \lambda}(i, j) + \sigma^2_n)\right),
\end{equation}
where the weighting factors are given by
\begin{equation}
   \alpha_{\lambda, \theta}(i, j) = \log\left(1 + \sigma^2_{\lambda, \theta}(i, j)\right).
\end{equation}
The difference map, i.e. spatial RRED (SRRED) quality map, is obtained as
\begin{equation}
    \mathrm{L\mhyphen SRRED}(i, j) = \left| h_{l, x, \lambda}(i, j) - h_{l, y, \lambda}(i, j)\right|
\end{equation}

Similarly, differences between local entropies of frame differences yield the temporal RRED (TRRED) quality map, and chroma SRRED and TRRED (C-SRRED and C-TRRED) maps were obtained by computing local entropies of the chroma channel under a complex Gaussian assumption. This procedure was repeated at four scales and for HDRMAX-transformed frames to yield a total of 32 RRED feature maps.

The final wavelet-domain feature corresponds to the Detail Loss Metric (DLM) \cite{ref:dlm}. A brief description of DLM is as follows. The first step in computing DLM involves applying a ``decoupling step.'' The decoupling step is based on the following distortion model of the wavelet subband coefficients \(\theta \in \{H, V, D\}\):
\begin{equation}
    Y_{\lambda, \theta}(i, j) = \gamma_{\lambda, \theta}(i, j) X_{\lambda, \theta}(i,j) + A_{\lambda, \theta}(i,j),
\end{equation}
where the gain factor \(\gamma\) models attenuation of local gradients due to detail loss, \(R_{\lambda, \theta}(i,j) = \gamma_{\lambda, \theta}(i, j) X_{\lambda, 1}(i,j)\) are the ``restored'' coefficients, and \(A_{\lambda, \theta}(i,j)\) are the ``additive impairments.''

The ``restored'' wavelet decomposition are computed from the given frames via
\begin{equation}
    \hat{R}_{\lambda, \theta}(i,j) = \hat{\gamma}_{\lambda, \theta}(i,j) X_{\lambda, \theta}(i,j),
\end{equation}
where
\begin{equation}
    \psi_{x, \lambda}(i,j) = \arctan\left(\frac{X_{\lambda, V}(i,j)}{X_{\lambda, H}(i,j)}\right),
\end{equation}
\begin{equation}
    \psi_{y, \lambda}(i,j) = \arctan\left(\frac{Y_{\lambda, V}(i,j)}{Y_{\lambda, H}(i,j)}\right),
\end{equation}
\begin{equation}
    \Delta\psi_{\lambda}(i, j) = \left|\psi_{x, \lambda}(i, j) - \psi_{y, \lambda}(i, j) \right|,
\end{equation}
and 
\begin{equation}
    \hat{\gamma}_{\lambda, \theta}(i, j) = \begin{cases}
        \frac{Y_{\lambda, \theta}(i,j)}{X_{\lambda, \theta}(i,j)}, & \Delta\psi_{\lambda}(i, j) < 1^{\circ} \\
        \text{clip}\left(\frac{Y_{\lambda, \theta}(i,j)}{X_{\lambda, \theta}(i,j)}, 0, 1\right), & \textit{else}
    \end{cases}.
\end{equation}

The quantity \(\Delta\psi\) is used to preserve contrast enhancement, which scales both the horizontal and vertical subband coefficients of the test frame \cite{ref:dlm}. The additive impairment coefficients are then computed as:
\begin{equation}
    \hat{A}_{\lambda, \theta}(i,j) = Y_{\lambda, \theta}(i,j) - \hat{R}_{\lambda, \theta}(i,j).
\end{equation}

The additive impairments are used to mask the restored coefficients using the model
\begin{equation}
    \Tilde{R}_{\lambda, \theta}(i,j) = \left(\hat{R}_{\lambda, \theta}(i,j) - M_{\lambda}(i,j)\right)^+,
\end{equation}
where
\begin{equation}
    M_{\lambda}(i,j) = \sum_{\theta}\sum\limits_{i-1}^{i+1}\sum\limits_{j-1}^{j+1} w_{ij}(m,n)\left|\hat{A}_{\lambda, \theta}(m,n)\right|,
\end{equation}
\begin{equation}
    w_{ij}(m,n) = \frac{1 + \delta(m-i, n-j)}{30},
\end{equation}
\((\cdot)^+\) denotes clipping negative values to zero, and \(\delta\) is the Kronecker delta function. Finally, modifying the definition in \cite{ref:dlm} to yield local scores, local DLM scores are computed as the following ratio:
\begin{equation}
    \mathrm{DLM}(i, j) = \frac{\left(\sum\limits_{\theta \in \{H, V, D\}} \Tilde{R}_{\lambda, \theta}(i, j)^{3}\right)^{1/3}}{\left(\sum\limits_{\theta \in \{H, V, D\}} \Tilde{X}_{\lambda, \theta}(i, j)^{3}\right)^{1/3}}.
\end{equation}
\subsection{Natural Scene Statistics Features}
\label{sec:cut_funque_nss_feature_extraction}
In addition to wavelet-domain features, we incorporated principles from natural scene statistics to compare input frames by their bandpass distributions. We achieved this by applying a bandpass transform called Mean Subtracted Contrast Normalization (MSCN), which has been used in a wide variety of popular NR quality prediction models, such as NIQE \cite{ref:niqe}, BRISQUE \cite{ref:brisque}, and HIGRADE \cite{ref:higrade}.

The MSCN transform is a Divisive Normalization Transform (DNT) applied to either the mean-subtracted luma or chroma channel
\begin{equation}
    \Tilde{I}(i, j) = \frac{I(i, j) - \mu(i, j)}{\sigma(i, j) + \epsilon},
\end{equation}
where \(\mu(i,j)\) and \(\sigma(i,j)\) are the local and means and standard deviations. Furthermore, to model contrast naturalness, we also applied the MSCN transform to the \(\sigma(i,j)\) field, yielding ``\(\sigma\)-MSCN'' coefficients. 

MSCN coefficients obtained in this manner have been observed to follow regular quality-aware statistical properties, which are captured by modeling their distribution using Generalized Gaussian Distributions (GGDs) \cite{ref:brisque} \cite{ref:higrade}. The GGD centered at 0 may be defined using its probability density function (PDF) as:
\begin{equation}
    f_{GGD}(x; \alpha, b) = \frac{\alpha}{2b\Gamma\left(\frac{1}{\alpha}\right)} \exp\left(-\left(\frac{|x|}{b}\right)^\alpha\right),
\end{equation}
where \(\alpha\) is called the ``shape parameter'', \(b\) is called the ``scale parameter'', and \(\Gamma\) denotes the Gamma function, which is defined as follows:
\begin{equation}
    \Gamma(x) = \int_{0}^{\infty} t^{x-1} e^{-t} dt.
\end{equation}

We fit GGDs to MSCN and \(\sigma-\)MSCN coefficients by estimating distribution parameters using a moment-matching procedure \cite{ref:ggd_est}, then use them as ``global'' no-reference NSS features. In addition, we also capture the second-order statistics of MSCN coefficients by modeling the distribution of products of spatially adjacent MSCN coefficients. We chose the neighboring pixel locations along four directions - horizontal, vertical, and two diagonals (\(H\), \(V\), \(D_1\), \(D_2\)). As in \cite{ref:brisque} and \cite{ref:niqe}, Asymmetric GGDs (AGGDs) are fit to the distributions of these products using a moment-matching approach \cite{ref:aggd_est} and their parameters used as quality-aware features. An AGGD centered at zero may be described by a PDF of the form
\begin{equation}
    f_{AGGD}(x; \alpha, b_l, b_r) = \begin{cases}
        \frac{\alpha}{\left(b_l + b_r\right)\Gamma\left(\frac{1}{\alpha}\right)} \exp\left(-\left(\frac{|x|}{b_l}\right)^\alpha\right) & x < 0, \\
        \frac{\alpha}{\left(b_l + b_r\right)\Gamma\left(\frac{1}{\alpha}\right)} \exp\left(-\left(\frac{|x|}{b_r}\right)^\alpha\right) & x \geq 0,
    \end{cases}
\end{equation}
where \(\alpha\) is a ``shape'' parameter and \(b_l, b_r\) are left and right ``scale parameters.''

To limit the number of features, we average the quality-aware AGGD parameter estimates over the four directions to obtain a final set of second-order global NR NSS features. 

The use of statistical fits to MSCN coefficients has typically been limited to the domain of NR quality assessment. However, the availability of a reference video opens up the possibility of defining ``statistical similarity'' (StSim) quality models, analogous to structural similarity models such as SSIM. Here, we introduce the novel method of comparing KL divergences between NSS distributions as a measure of quality-aware statistical dissimilarity. 

Let a pair of corresponding GGD parameters obtained from a reference (HDR) and distorted (SDR) frame pair be \((\alpha_1, b_1)\) and \((\alpha_2, b_2)\). Then, the first-order statistical dissimilarity (FOSD) is the KL Divergence between the two GGDs:
\begin{align}
    \mathrm{FOSD} = \frac{\Gamma\left(\frac{\alpha_2+1}{\alpha_1}\right)}{\Gamma(\frac{1}{\alpha_1})} \left(\frac{b_1}{b_2}\right)^{\alpha_2} - \frac{1}{\alpha_1} + \log\left(\frac{\alpha_1 b_2 \Gamma\left(\frac{1}{\alpha_2}\right)}{\alpha_2 b_1 \Gamma\left(\frac{1}{\alpha_1}\right)}\right).
\end{align}

Similarly, the second-order statistical dissimilarity (SOSD) between two frames described by AGGDs \((\alpha_1, b_{l1}, b_{r1})\) and \((\alpha_2, b_{l2}, b_{r2})\) is:
\begin{align}
    \mathrm{SOSD} = &\frac{\Gamma\left(\frac{\alpha_2 + 1}{\alpha_1}\right)}{\Gamma\left(\frac{1}{\alpha_1}\right)\left(b_{l1}+b_{r1}\right)}\left(\frac{b_{l1}^{\alpha_2+1}}{b_{l2}^{\alpha_2}} + \frac{b_{r1}^{\alpha_2+1}}{b_{r2}^{\alpha_2}}\right) - \frac{1}{\alpha_1} + \nonumber \\&\log\left(\frac{\alpha_1\left(b_{l2} + b_{r2}\right)\Gamma\left(\frac{1}{\alpha_2}\right)}{\alpha_2\left(b_{l1} + b_{r1}\right)\Gamma\left(\frac{1}{\alpha_1}\right)}\right).
\end{align}

We repeated the same procedure locally to obtain first and second-order NSS fits to coefficients within cuts of size \(8\times 8\), \(16\times16\), \(32\times32\), and \(64 \times 64\), corresponding to analyzing the video at four scales. Then, FOSD and SOSD estimates were obtained for each cut, yielding local statistical dissimilarity estimates.
\subsection{Binned Weighting to Isolate Frame Regions}
\label{sec:cut_funque_weighted_bins}
As explained earlier, we expect different regions of a frame that vary in their characteristics to be affected differently by tone-mapping and compression distortions. So, we first partition the frame into ``cuts'', i.e., non-overlapping patches of equal sizes to enable local analysis. CutQA utilizes a multi-scale approach, with the finest scale corresponding to cuts of size \(8\times 8\) and the coarsest scale corresponding to \(64\times 64\) cuts.

Next, in accordance with the three observations made earlier, we characterize the \textbf{luminance} (brightness), \textbf{spatial complexity}, and \textbf{temporal complexity} of each cut using three low-level measures. The brightness of the cut is characterized by the mean luma value over the cut. Spatial complexity, i.e., contrast, is often characterized by the standard deviation of luma values within the cut. However, brighter patches tend to have higher standard deviations since their pixel values are greater. To account for this phenomenon, we instead used the coefficient-of-variation (CoV), defined as the ratio of standard deviation to mean. Finally, to characterize the temporal complexity, we computed the averaged temporal coefficient of variation of luma values over the previous four frames.

Formally, consider a cut \(c\) and an input HDR frame \(I_{HDR}\), the three low-level measures are computed as

\begin{equation}
L(c) = E_{(i,j) \in c}\left[I_{HDR}\left(i,j\right)\right],
\end{equation}

\begin{equation}
S(c) = \frac{Std_{(i,j) \in c}\left[I_{HDR}\left(i,j\right)\right]}{E_{(i,j) \in c}\left[I_{HDR}\left(i,j\right)\right]},
\end{equation}
and
\begin{equation}
T(c) = \frac{E_{(i,j) \in c}\left[Std_{t \in [-3, 0]}\left[I_{HDR}\left(i,j\right)\right]\right]}{E_{(i,j) \in c}\left[E_{t \in [-3, 0]}\left[I_{HDR}\left(i,j\right)\right]\right]},
\end{equation}
where \(E_{(i,j) \in c}\) and \(Std_{(i,j) \in c}\) denote computing means and standard deviations over spatial locations in a cut \(c\), and \(E_{t \in [-3, 0]}\) and \(Std_{t \in [-3, 0]}\) denote computing mean and standard deviation over the four most recent frames, including the current frame. Local means and standard deviations used to compute \(L(c)\) and \(S(c)\) are computed from the perceptually-sensitized Haar wavelet decomposition described in Section \ref{sec:cut_funque_wavelet_transform}. The four most recent frames used to compute \(T(c)\) are stored in a ``reference frame buffer,'' which is updated framewise.

The set of low-level descriptors obtained in this manner are then assigned to four equally spaced bins of each type, respectively called luminance, spatial, and temporal contrast (L-, S-, and T-) bins. In particular, for each bin type, a Gaussian membership function is used to define four membership weights for each cut. For example, consider a cut \(c\) with mean luminance \(L(c)\), and let the four luminance bins have centers \(\Tilde{L}_b\) and widths \(w^{(L)}\) (\(b = 0, 1, 2, 3\)). The membership weights for the cut corresponding to each luminance bin are computed as
\begin{equation}
    M^{(L)}_b(c) = \exp\left(-\frac{\left(L(c) - \Tilde{L}_b\right)^2}{2\left(w^{(L)}/2\right)^2}\right).
    \label{eq:bin_weight}
\end{equation}
This yields a soft classification of the cut into four luminance bins. Using the same procedure, each cut is also soft classified into four S-bins and four T-bins using weights \(M^{(S)}_b(c)\) and \(M^{(S)}_b(c)\) respectively, which are calculated as in Eq. \eqref{eq:bin_weight}.

\begin{figure*}[t]
    \centering
    \subfloat[width=0.25\linewidth][Input video frame]{\includegraphics[width=0.25\linewidth]{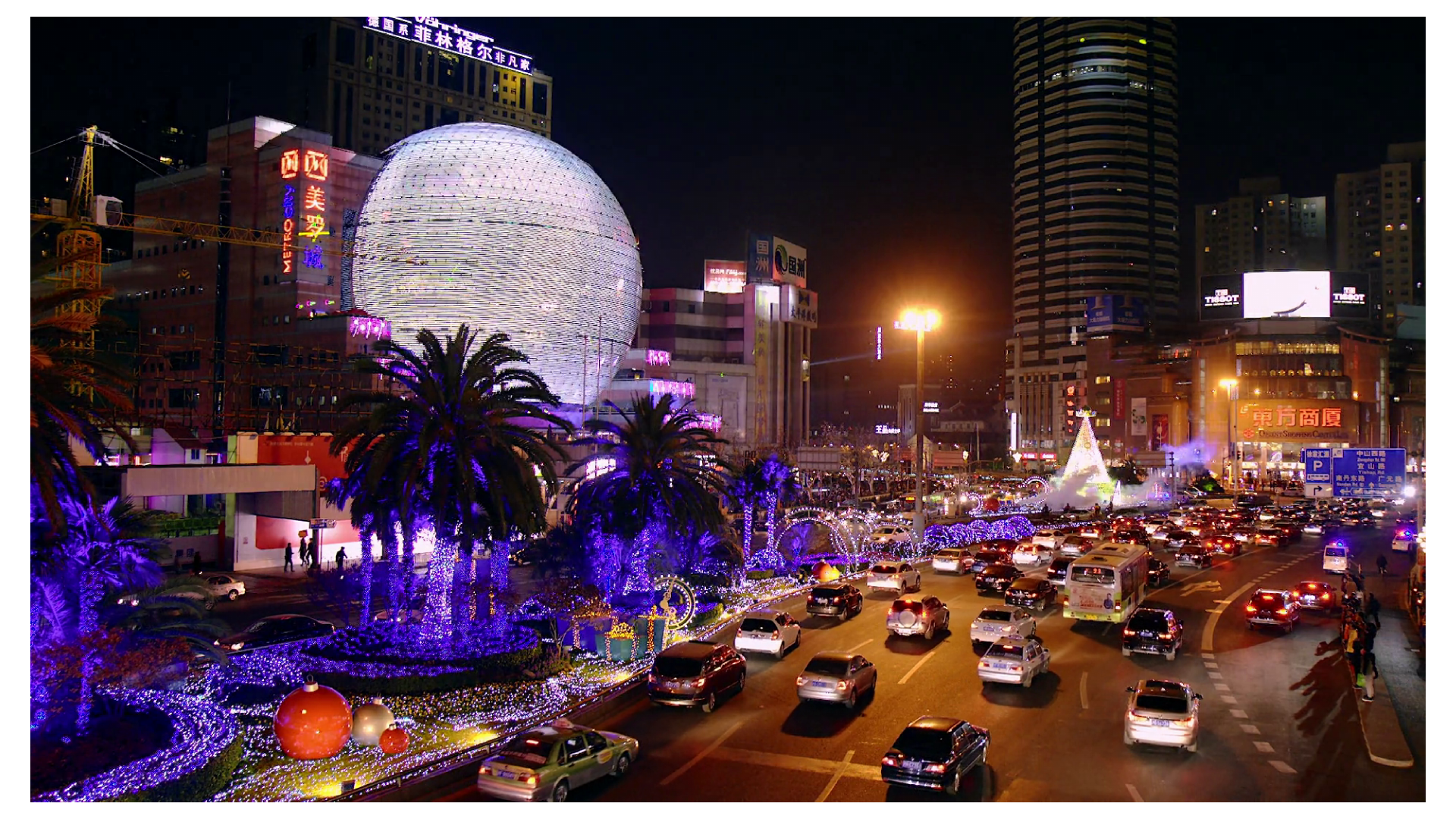}\label{fig:nightraffic_sample}}
    \\
    \subfloat[width=0.25\linewidth][Luminance bin 0]{\includegraphics[width=0.25\linewidth]{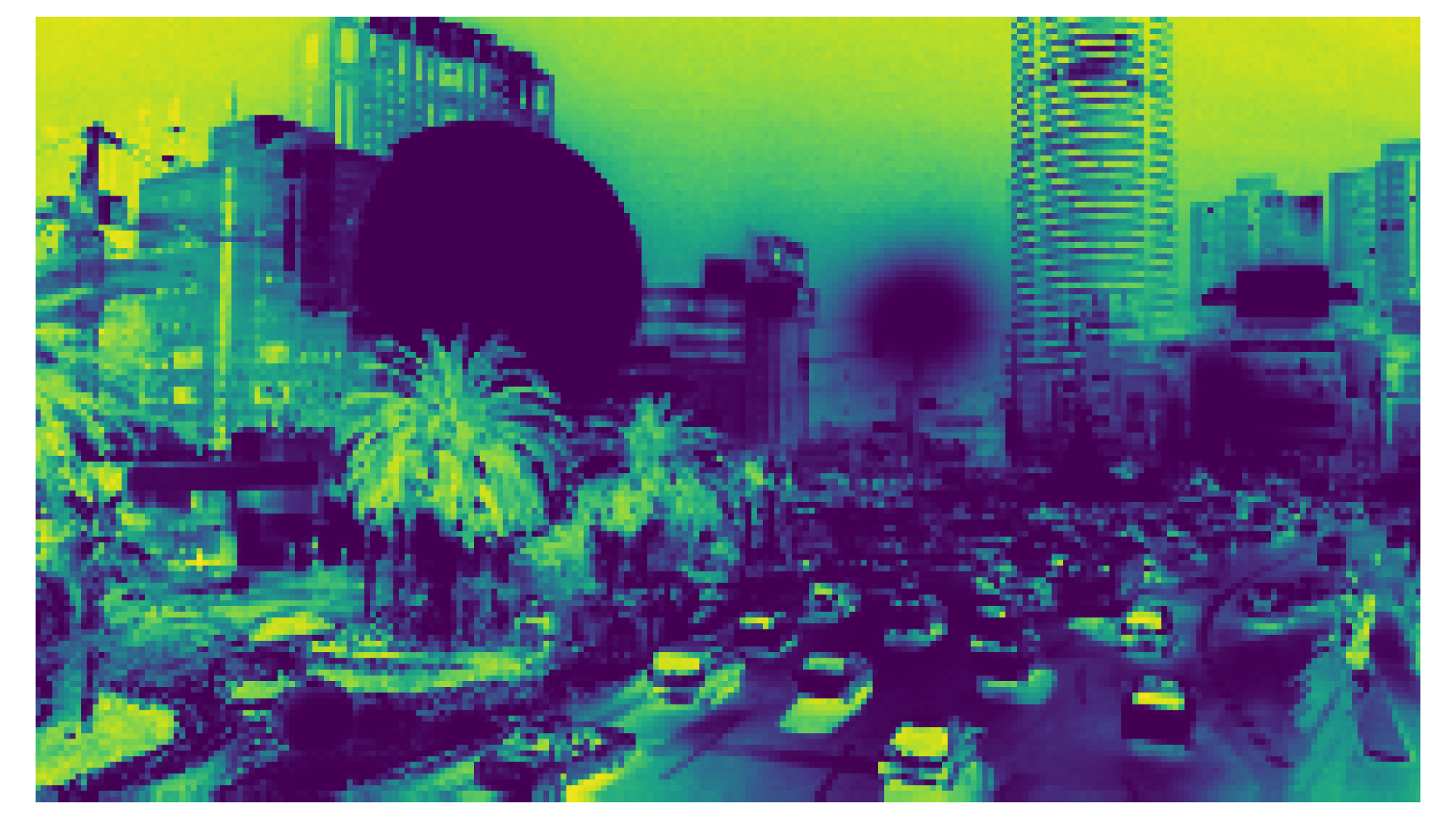}\label{fig:nightraffic_cut_weights_lum_bin_0}}%
    \subfloat[width=0.25\linewidth][Luminance bin 1]{\includegraphics[width=0.25\linewidth]{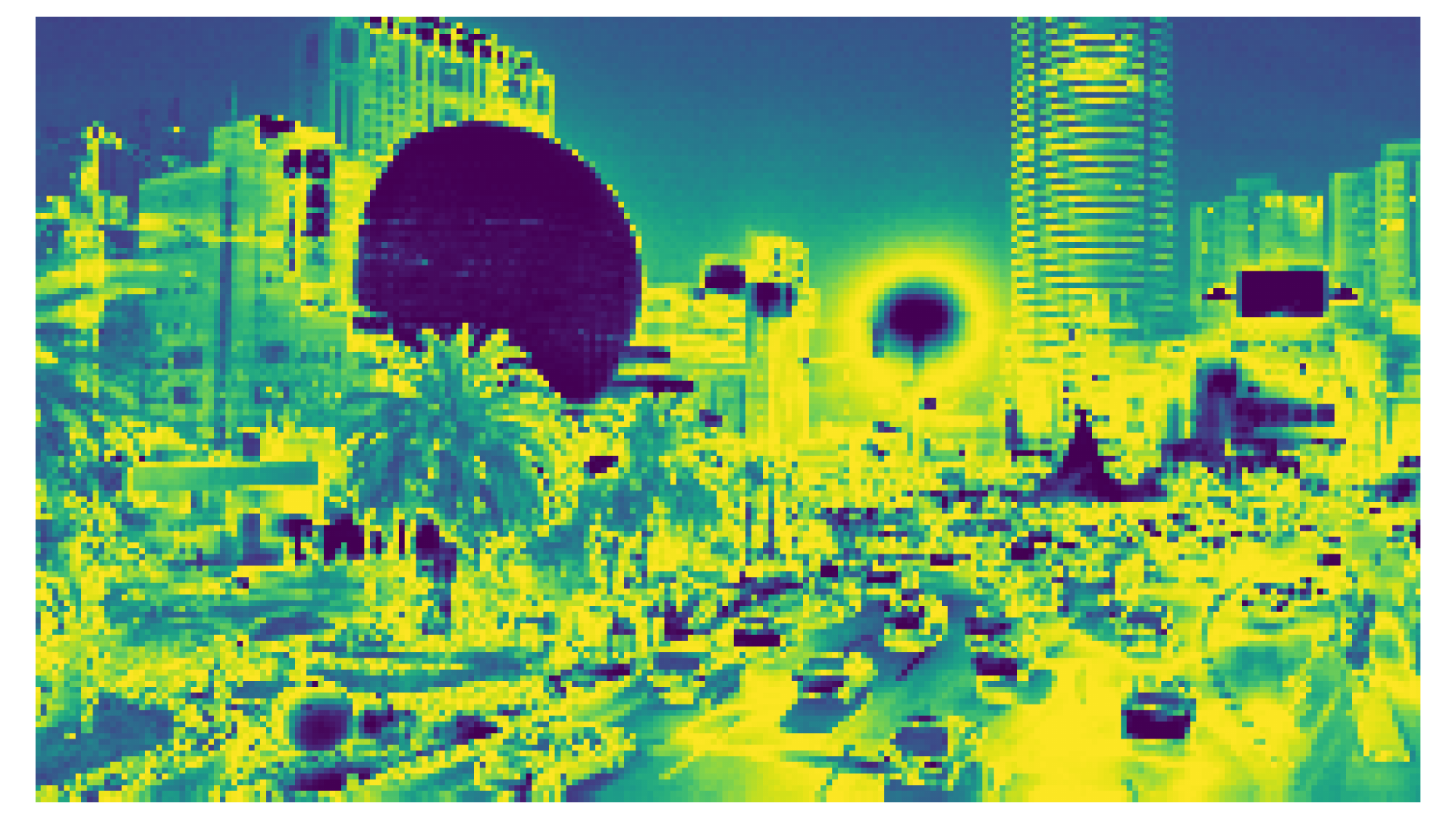}\label{fig:nightraffic_cut_weights_lum_bin_1}}%
    \subfloat[width=0.25\linewidth][Luminance bin 2]{\includegraphics[width=0.25\linewidth]{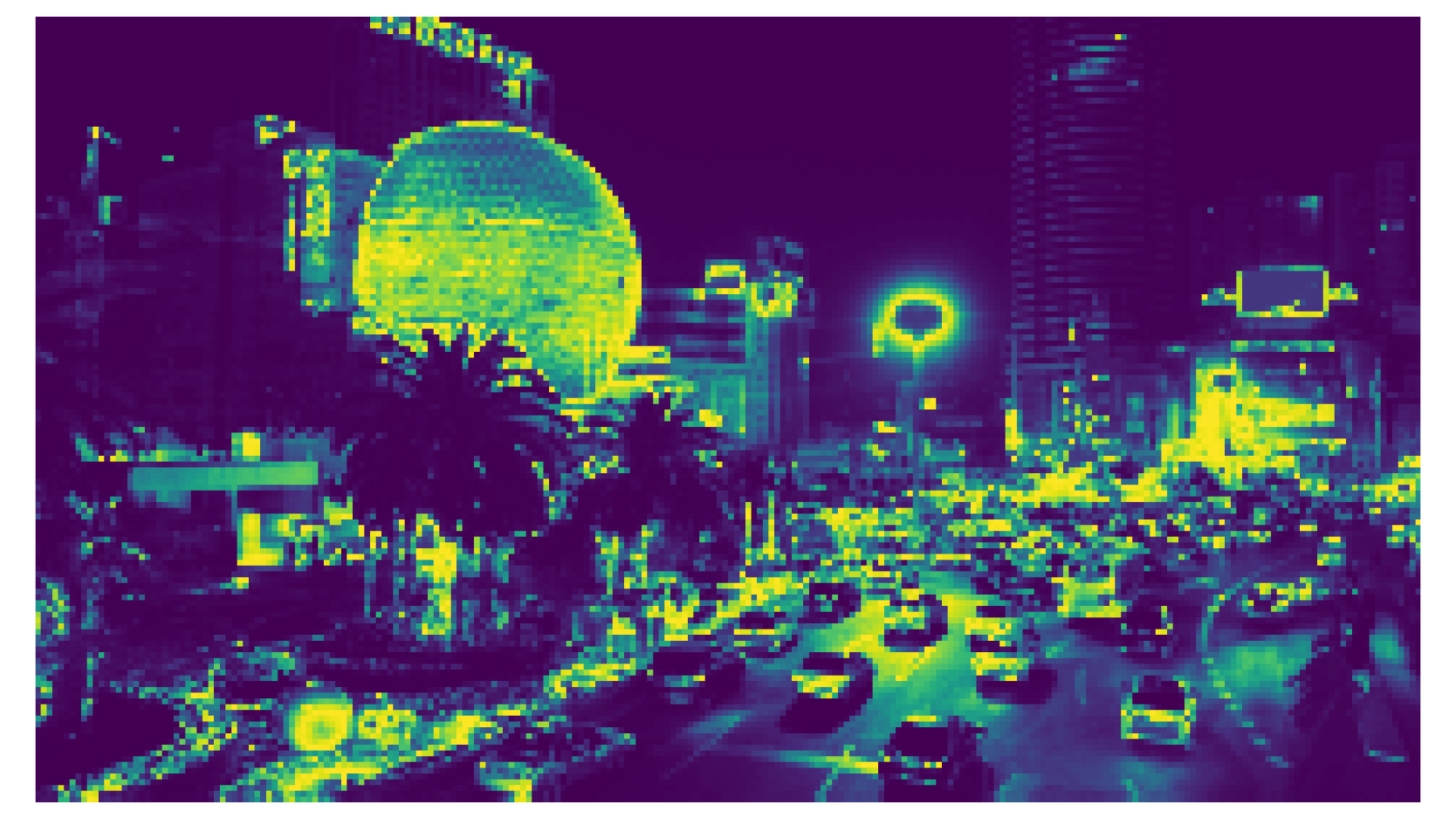}\label{fig:nightraffic_cut_weights_lum_bin_2}}%
    \subfloat[width=0.25\linewidth][Luminance bin 3]{\includegraphics[width=0.25\linewidth]{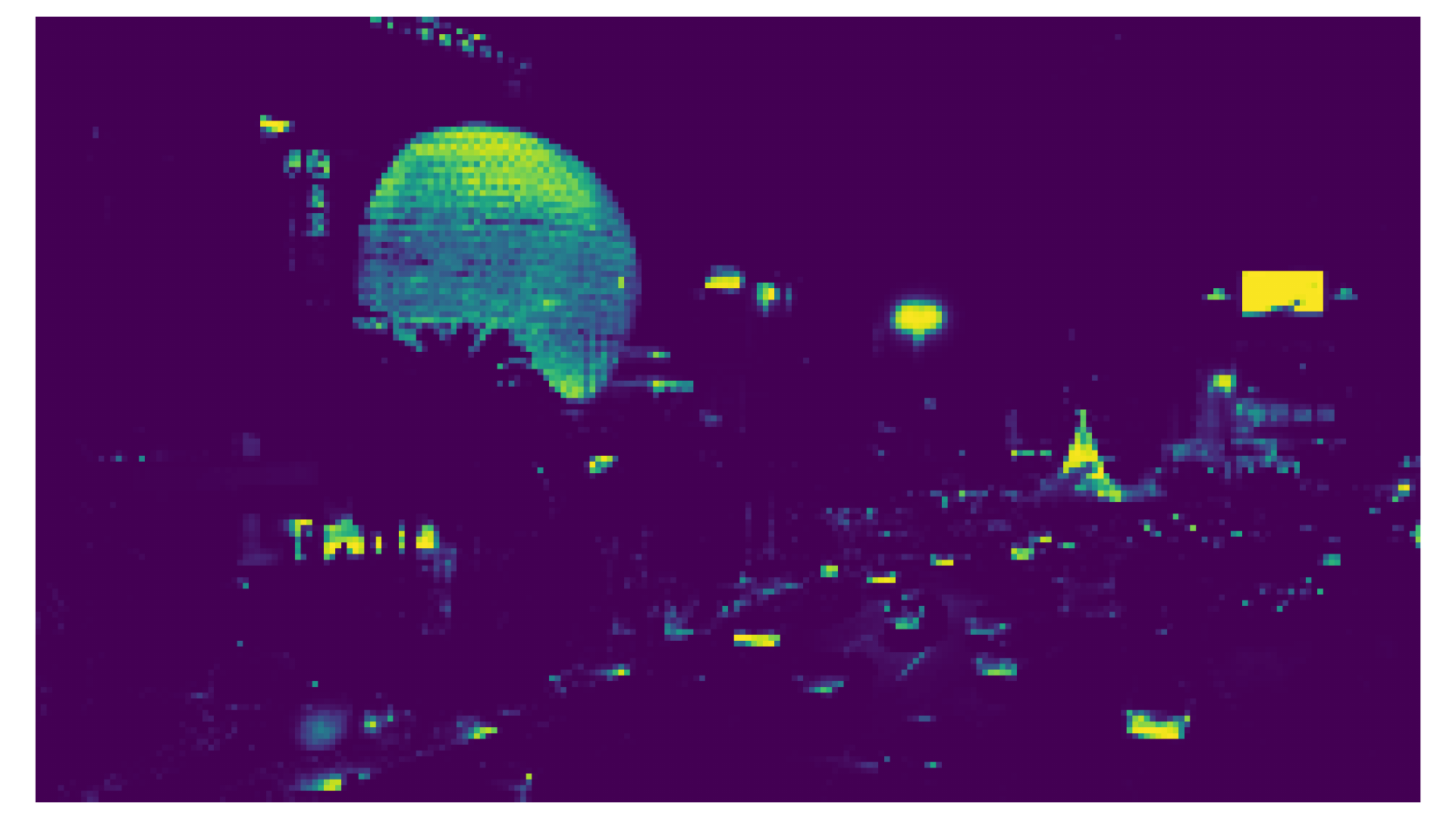}\label{fig:nightraffic_cut_weights_lum_bin_3}}%
    \\
    \subfloat[width=0.25\linewidth][Spatial complexity bin 0]{\includegraphics[width=0.25\linewidth]{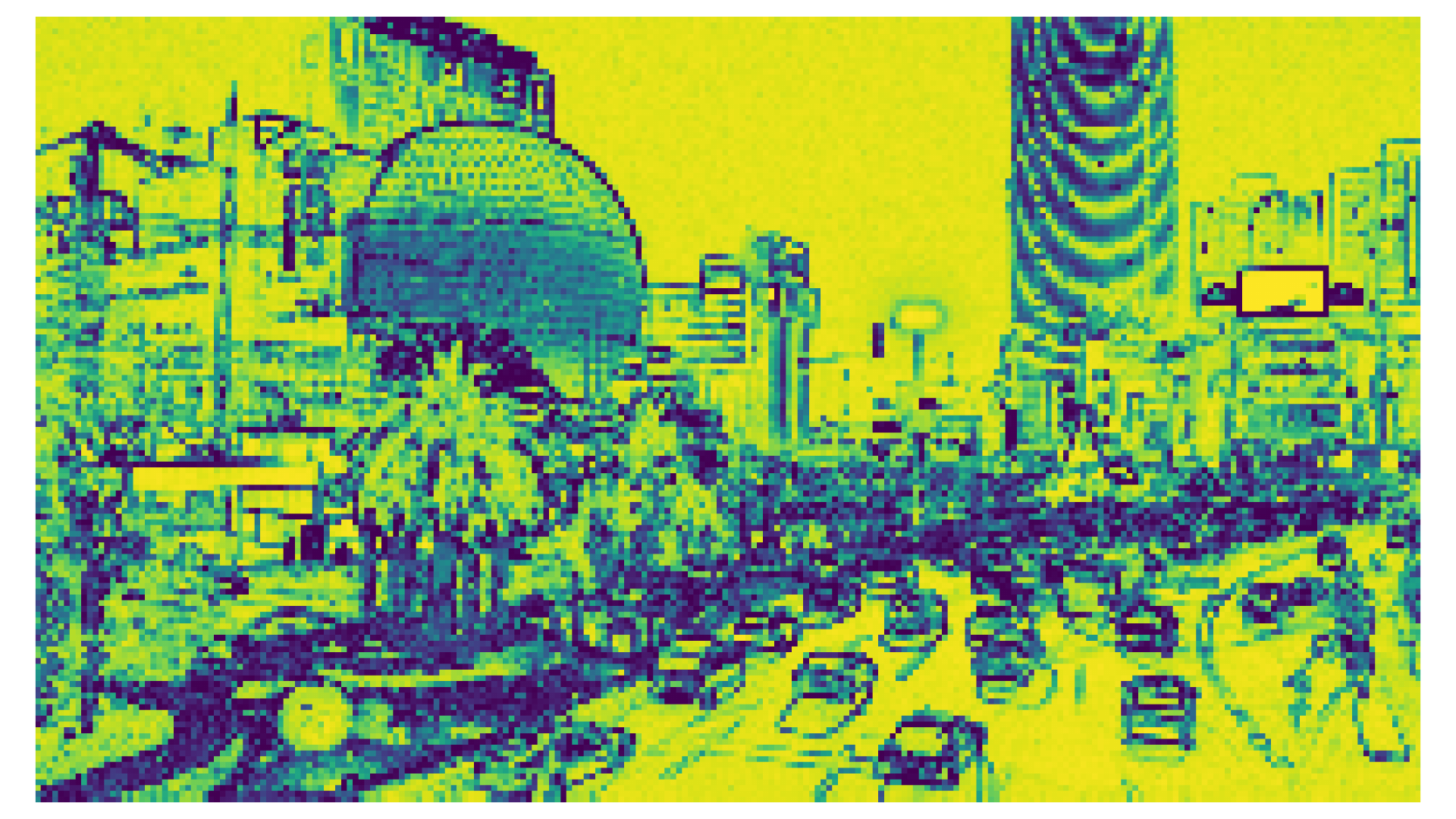}\label{fig:nightraffic_cut_weights_spat_bin_0}}%
    \subfloat[width=0.25\linewidth][Spatial complexity bin 1]{\includegraphics[width=0.25\linewidth]{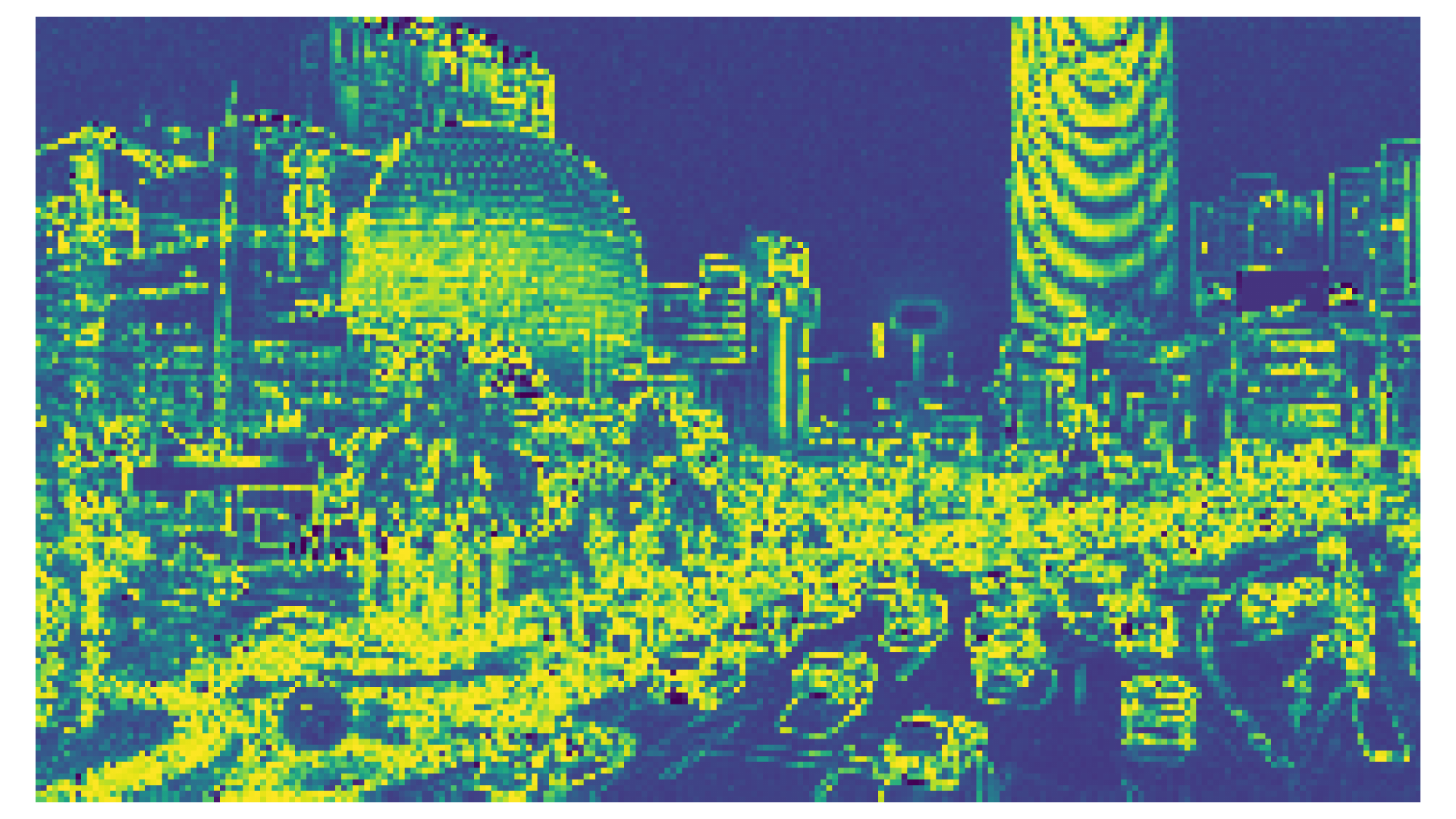}\label{fig:nightraffic_cut_weights_spat_bin_1}}%
    \subfloat[width=0.25\linewidth][Spatial complexity bin 2]{\includegraphics[width=0.25\linewidth]{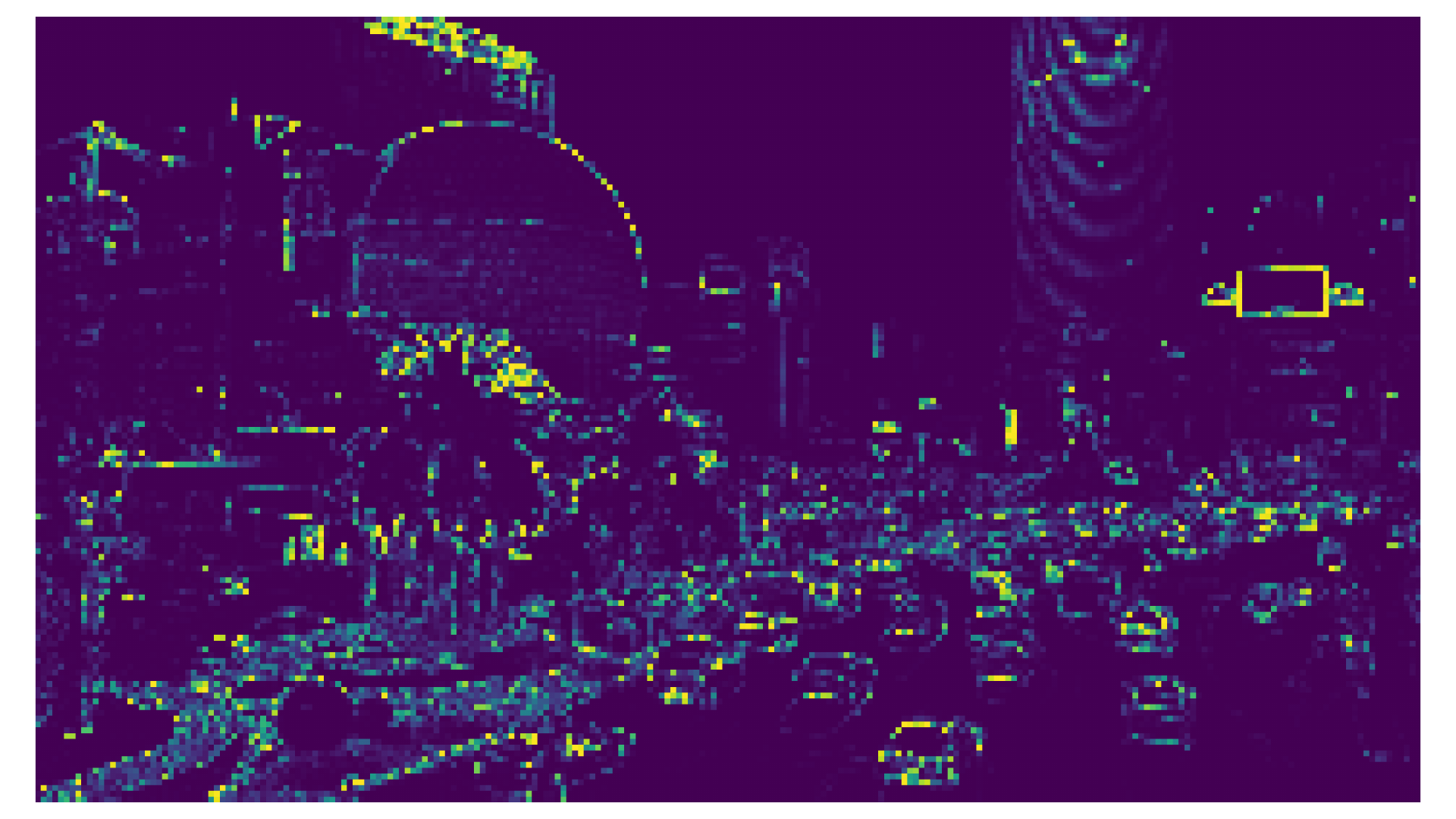}\label{fig:nightraffic_cut_weights_spat_bin_2}}%
    \subfloat[width=0.25\linewidth][Spatial complexity bin 3]{\includegraphics[width=0.25\linewidth]{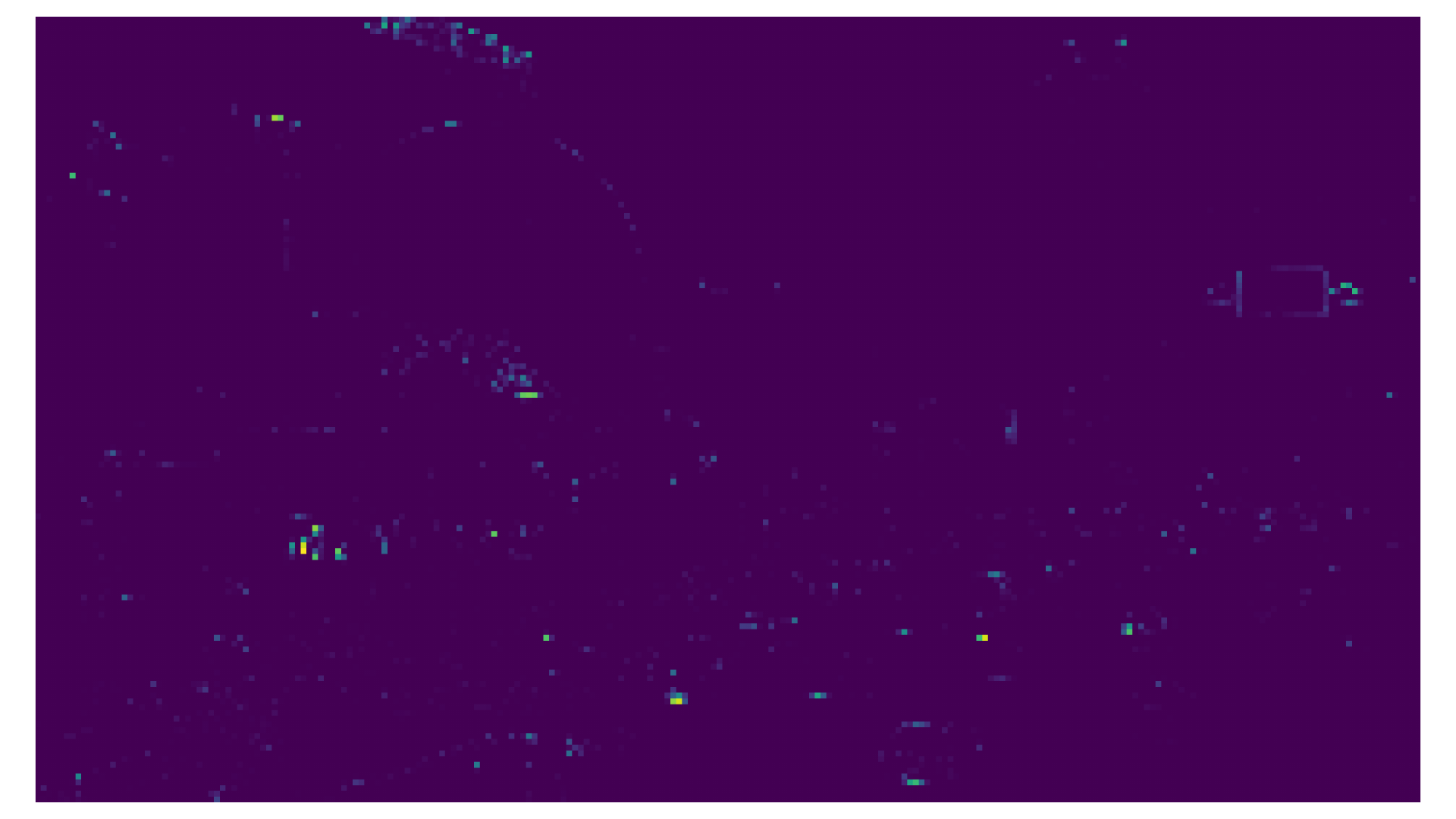}\label{fig:nightraffic_cut_weights_spat_bin_3}}%
    \\
    \subfloat[width=0.25\linewidth][Temporal complexity bin 0]{\includegraphics[width=0.25\linewidth]{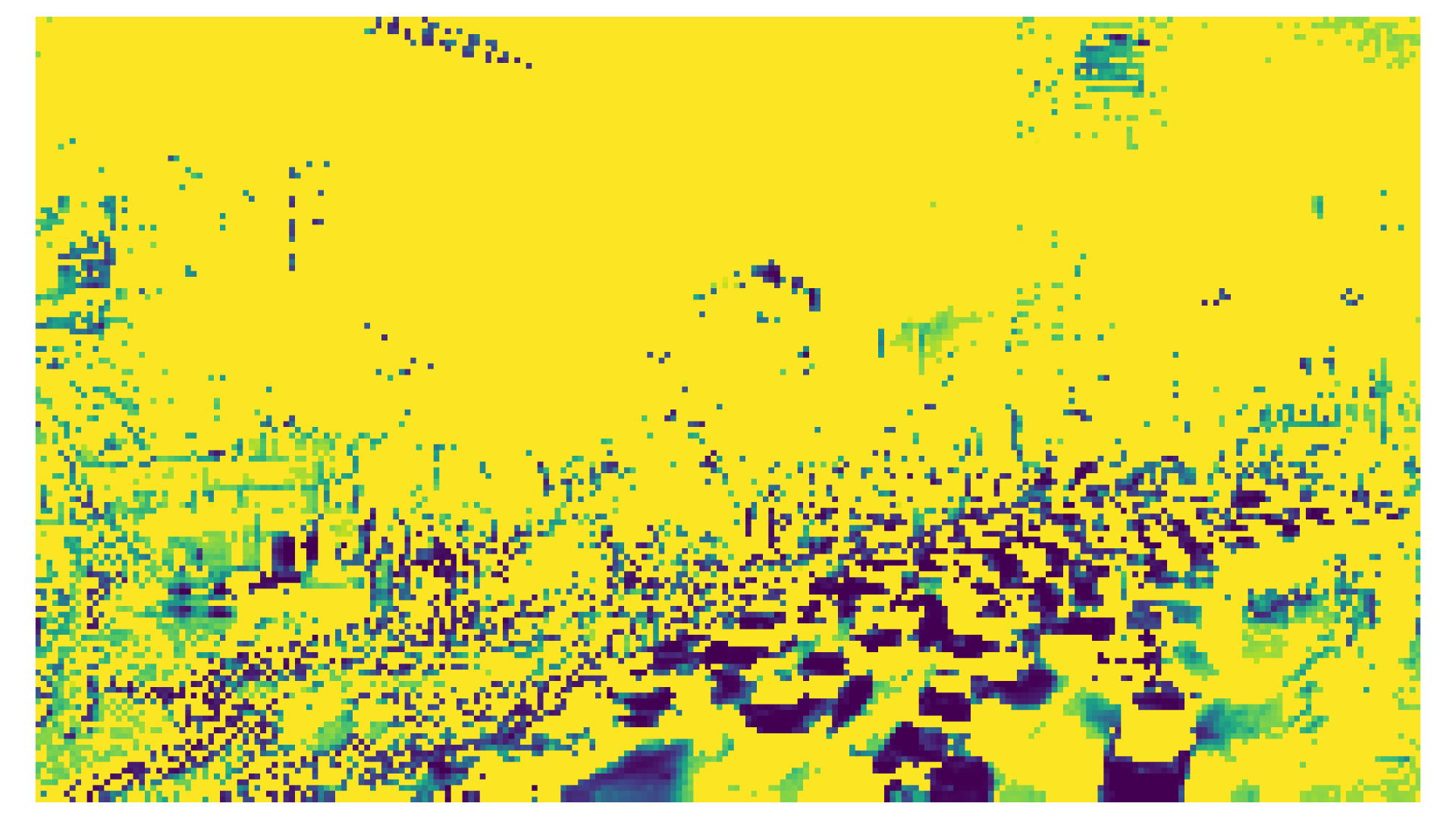}\label{fig:nightraffic_cut_weights_temp_bin_0}}%
    \subfloat[width=0.25\linewidth][Temporal complexity bin 1]{\includegraphics[width=0.25\linewidth]{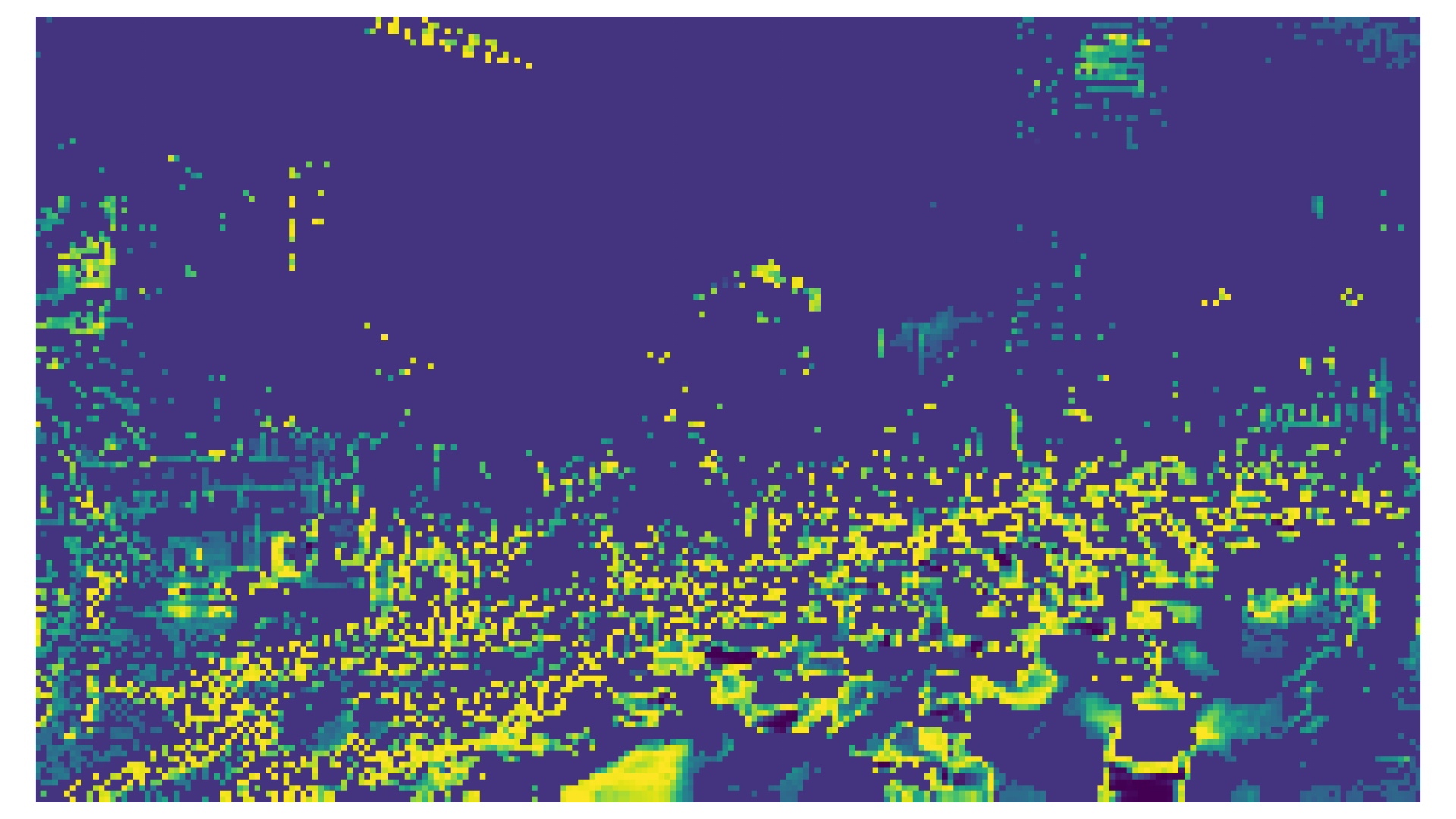}\label{fig:nightraffic_cut_weights_temp_bin_1}}%
    \subfloat[width=0.25\linewidth][Temporal complexity bin 2]{\includegraphics[width=0.25\linewidth]{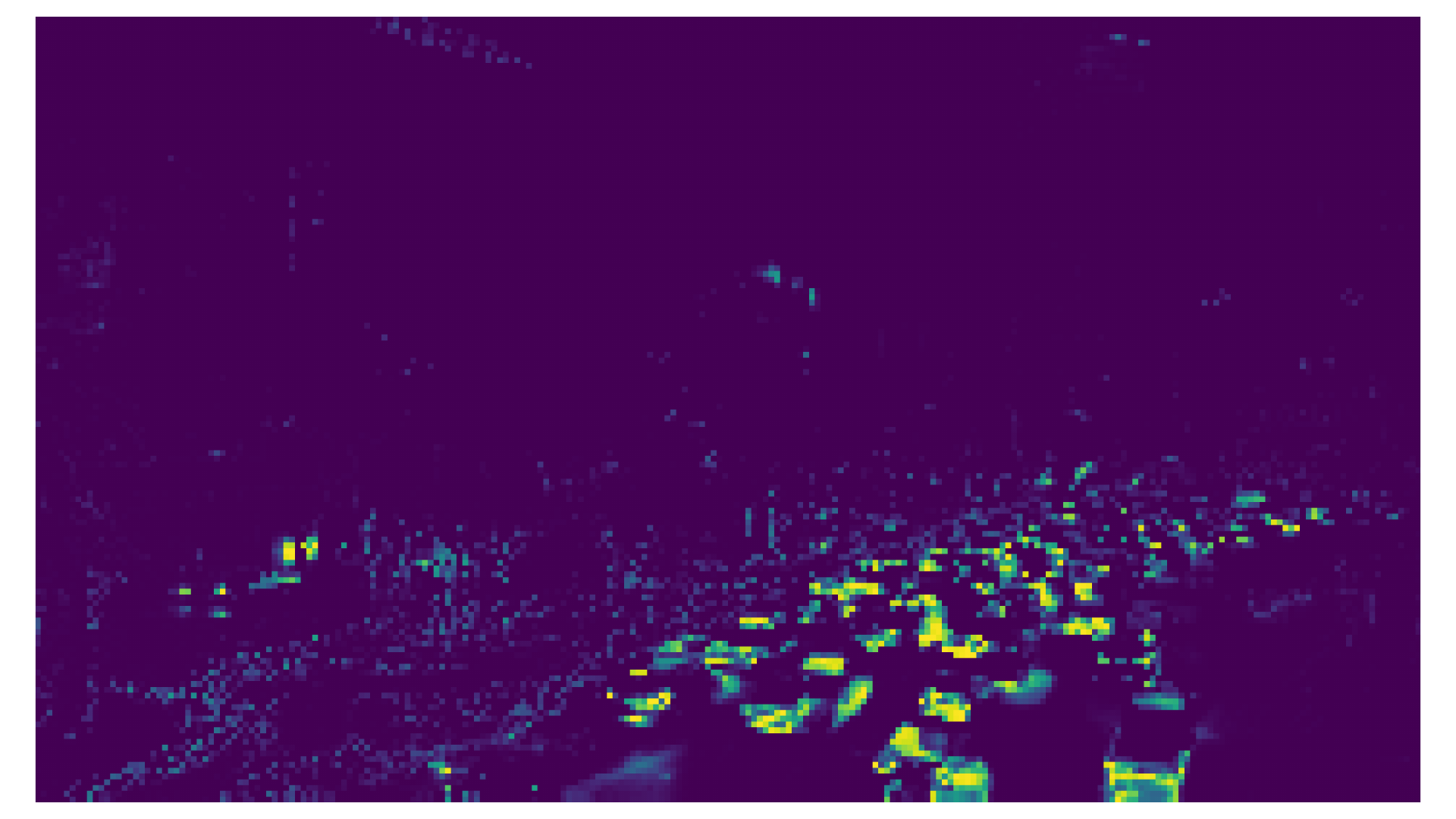}\label{fig:nightraffic_cut_weights_temp_bin_2}}%
    \subfloat[width=0.25\linewidth][Temporal complexity bin 3]{\includegraphics[width=0.25\linewidth]{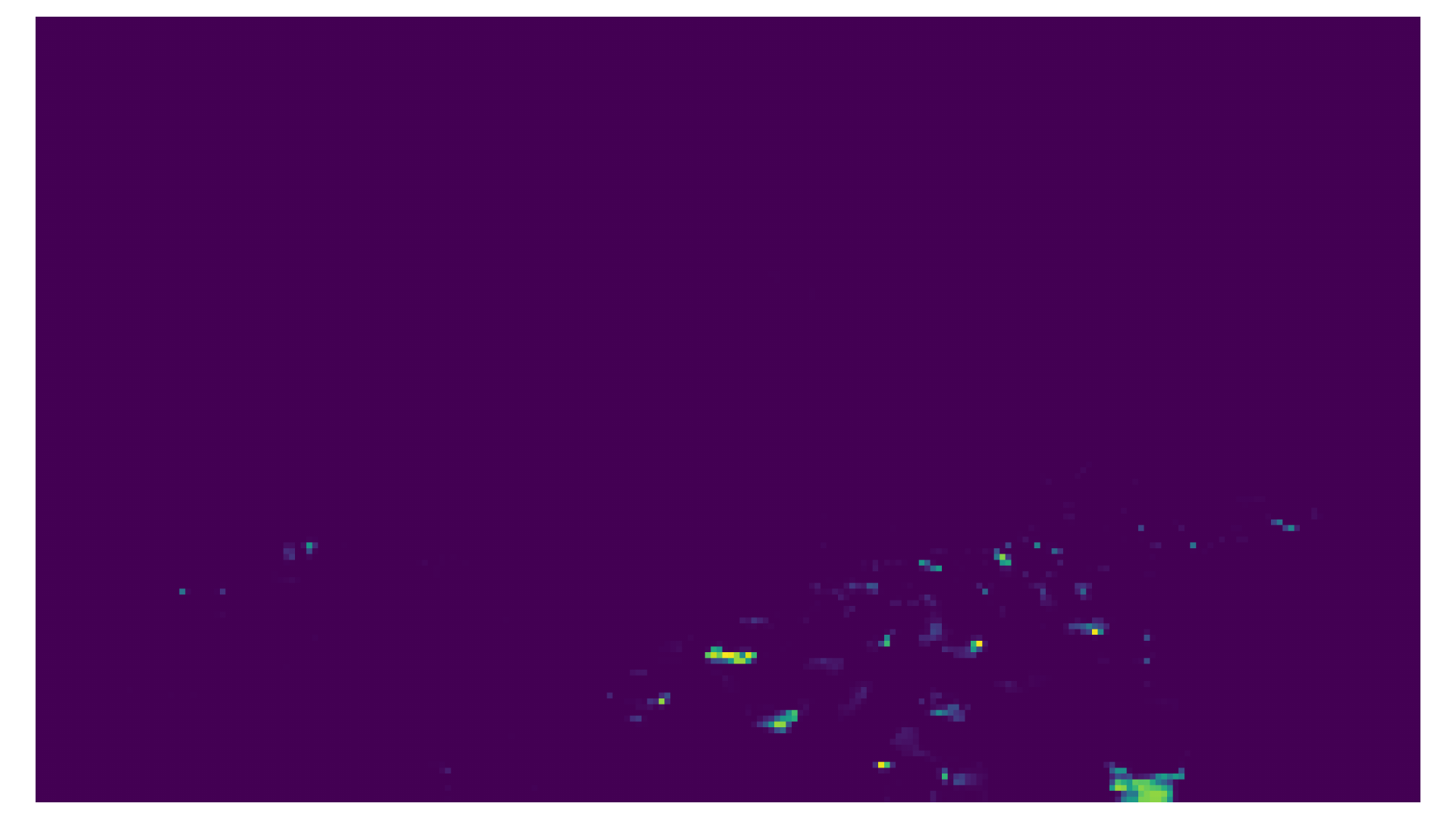}\label{fig:nightraffic_cut_weights_temp_bin_3}}%
\caption{A sample video frame and the soft-classification of \(8\times8\) cuts into four bins based on luminance, spatial, and temporal complexity features.}
\label{fig:uniform_cuts}
\end{figure*}

An example of partitioning a video frame into \(8\times 8\) cuts and corresponding bin classification weights is presented in Fig. \ref{fig:uniform_cuts}. The scene (Fig. \ref{fig:nightraffic_sample}) consists of cars moving on a road surrounded by various buildings and lights. Figs. \ref{fig:nightraffic_cut_weights_lum_bin_0}-\ref{fig:nightraffic_cut_weights_lum_bin_3} visualize L-bin membership weights \(M^{(L)}_0\)-\(M^{L}_3\), Figs. \ref{fig:nightraffic_cut_weights_spat_bin_0}-\ref{fig:nightraffic_cut_weights_spat_bin_3} visualize S-bin membership weights \(M^{(S)}_0\)-\(M^{S}_3\), and Figs. \ref{fig:nightraffic_cut_weights_temp_bin_0}-\ref{fig:nightraffic_cut_weights_temp_bin_3} visualize T-bin membership weights \(M^{(T)}_0\)-\(M^{T}_3\). 

In each weight map, a brighter yellow color denotes a higher membership weight, i.e., the corresponding region ``belongs more'' to the bin. From that figure, it may be observed that the luminance bins distinguish between bright and dark regions, while spatial complexity bins distinguish between detailed regions, and temporal complexity bins can segment moving objects such as cars.

Bin-membership weights are then used to aggregate local quality scores to bin-level quality scores. Bin-level quality scores represent the quality of different ``types of regions'' in the image. For example, quality scores aggregated within luminance bin 0 encode the quality of dark image regions. These scores are obtained using weighted averaging, which we illustrate using the example of SRRED. Let \(SRRED(c)\) denote the SRRED feature value computed for each cut \(c\). The weighted-aggregated SRRED values for the four luminance bins (\(b = 0, 1, 2, 3\)) are computed as
\begin{equation}
    SRRED^{(L)}_b = \frac{\sum_{c} M^{(L)}_b(c) \times SRRED(c)} {\sum_{c} M^{(L)}_b(c)}.
\end{equation}
Similarly, \(SRRED^{(S)}_b\) and \(SRRED^{(T)}_{b}\) are computed, and the process is repeated for all features. 

The one exception is SSIM, which is aggregated using weighted Minkowski pooling, based on the recommendations in \cite{ref:essim}. We preferred Minkowski pooling over the CoV pooling used in \cite{ref:funque} because we found the ratio used in CoV pooling to be numerically unstable for low quality inputs. Hence, SSIM is aggregated as
\begin{equation}
    SSIM^{(L)}_b = 1 - \left(\frac{\sum_{c} M^{(L)}_b(c) \times \left(1 - SSIM(c)\right)^3} {\sum_{c} M^{(L)}_b(c)}\right)^{1/3}.
\end{equation}

Prior work on spatial quality aggregation \cite{ref:essim} has revealed that low-quality spatial regions are more salient to observers and, therefore, largely determine the overall perceived quality of the input stimulus. In line with this observation, we posit that the most salient bin of each type, which corresponds to spatial regions in the image, is that which has the lowest quality. So, we aggregate bin-level quality features into a single feature per scale by using the quality of the worst-quality bin. For quality features such as SSIM and VIF (i.e., a greater value implies better quality), this involves applying a \(\min\) operation over bin-level features, while for distortion features such as SRRED and FOSD (i.e, a greater value implies worse quality), this involves applying a \(\max\) operation over bin-level features.

Although perceptually motivated, using only the quality features from the worst image regions ignores other regions of the input frame. So, we also obtain global estimates of quality-aware features at each scale by computing ``unweighted averages,'' i.e., the mean value without binning or weighting. These features also function as a baseline against which the efficacy of binned-weighting is evaluated. The results of detailed ablation studies studying the effect of weighted aggregation and the various bin types are presented in Section \ref{sec:evaluation_tm_results}.

The procedure described thus yields binned-weighted and unweighted quality features computed at four scales. Finally, these features are aggregated across the four scales to yield a single multi-scale feature using the multi-scale fusion weights proposed in MS-SSIM \cite{ref:ms_ssim}. Let \(w_s\) denote the MS-SSIM fusion weight for scale \(s\) and let a quality feature at that scale be \(Q_s\). Then, the multi-scale quality feature was computed as:
\begin{equation}
    Q = \frac{\sum_s w_s Q_s}{\sum_s w_s}.
\end{equation}

The combination of global NSS features described in Section \ref{sec:cut_funque_nss_feature_extraction} and the multi-scale features computed as described above yields a total of 232 features.
\section{Evaluating Objective Quality Models}
\label{sec:evaluation}
In this section, we report the results of evaluating the effectiveness of the Cut-FUNQUE Model against a set of baseline quality models from the literature. In addition, we analyzed various combinations of the novel proposals made in the Cut-FUNQUE model in the form of an ablation study. To conduct these evaluations, we utilized the recently developed LIVE-TMHDR database \cite{ref:live_tmhdr}.

\subsection{The LIVE-TMHDR Database}
\label{sec:evaluation_live_tmhdr}
The LIVE-TMHDR Database is a recently-developed first-of-its-kind large-scale subjective database containing a set of 40 source HDR videos, 20 of which were created by professional studios and 20 generated by amateur iPhone users. The set of 40 source contents were subjected to both tone-mapping, using a wide variety of algorithms and manually tone-mapped by an expert colorist, and video compression, using the x264 \cite{ref:x264} encoder, yielding a total of 15,000 test contents.

The new database represents twelve tone-mapping operators (TMOs), of which ten were made open-source. The ten open-source TMOs and their key features may be summarized as follows.
\begin{enumerate}
\item \textbf{Hable} \cite{ref:hable} - A parameter-free pointwise non-linear transform originally designed for use in the video game \textit{Uncharted 2}. 

\item \textbf{Reinhard02} \cite{ref:reinhard02} - A point non-linearity to map luminances from HDR to SDR.

\item \textbf{Durand02} \cite{ref:durand02} - Uses a ``fast bilateral filter'' to decompose the luminances of HDR frames into ``base'' and ``detail'' layers.

\item \textbf{Shan12} \cite{ref:shan12} - Uses an edge-aware stationary wavelet transform (SWT) \cite{ref:swt}.

\item \textbf{Reinhard12} \cite{ref:reinhard12} - Uses color-appearance models applied in a local manner.

\item \textbf{Eilertsen15} \cite{ref:eilertsen15} - Applies a ``fast detail extraction'' method to obtain a base-detail decomposition and applies a dynamic tone-curve.

\item \textbf{Oskarsson17} \cite{ref:oskarsson17} - Uses Dynamic Programming to cluster values in the input image channels.

\item \textbf{Rana19} \cite{ref:rana19} - Uses a Generative Adversarial Network (GAN) to create a fully-convolutional, parameter-free TMO.

\item \textbf{Yang21} \cite{ref:yang21} - Uses a deep convolutional neural network (CNN) to transform a multi-scale Laplacian pyramid decomposition of each input HDR frame.

\item \textbf{ITU21} \cite{ref:itu21} - A parameter-free TMO proposed by the ITU in Recommendation BT.2446 (``Approach A'').
\end{enumerate}

In addition to these methods, the DolbyVision TMO (DV), created by Dolby as part of the DolbyVision HDR standard, and the Color Space Transform (CST) provided in DaVinci Resolve, which is a popular gamut/tone-mapping tool used by colorists, were also included in the database. Finally, all videos were encoded at three compression levels using the libx264 \cite{ref:x264} encoder. Due to its large size and diversity of source contents and distortions, we chose LIVE-TMHDR as a test bench for evaluating quality prediction models aimed at the practical delivery of tone-mapped HDR videos\footnote{Experiments using the LIVE-TMHDR database were conducted at the University of Texas at Austin by university-affiliated authors.}.
\subsection{Evaluation Protocol}
\label{sec:evaluation_protocol}
To evaluate Cut-FUNQUE and compare it with quality prediction models from the literature, we conducted content-separated cross-validation and report the median Pearson Correlation Coefficient (PCC), Spearman's Rank Order Correlation Coefficient (SROCC), and Root Mean Square Error (RMSE) over 100 random splits of the LIVE-TMHDR database. 

In each split, 80\% of the data was used for training and 20\% was used for testing. When generating random splits, we ensured that test contents from the same source HDR video were not present in both the training and test splits. To show each quality prediction feature set in its best light, we experimented with three regression models (wherever feasible) - the Linear Support Vector Regressor (SVR), Gaussian SVR, and Random Forest Regressor - to map features to quality scores. Hyper-parameters of each regressor were also tuned using cross-validation, and the regressor that achieved the highest cross-validation accuracy was selected. In the case of MSML, the partial least-squares projection matrix was recalibrated on the training dataset of each cross-validation split for a fair evaluation.
\subsection{Results of Quality Prediction on LIVE-TMHDR}
\label{sec:evaluation_tm_results}

To illustrate the efficacy of Cut-FUNQUE, we evaluated its performance against 15 TM-HDR quality prediction models from the literature. TMQI \cite{ref:tmqi}, FSITM \cite{ref:fsitm}, and FFTMI \cite{ref:fftmi} are full-reference image quality assessment (FR IQA) models, TMVQI \cite{ref:tmvqi} and FUNQUE+ \cite{ref:funque_plus} are FR video quality assessment (FR VQA) models, and BRISQUE \cite{ref:brisque}, NIQE \cite{ref:niqe}, DIIVINE \cite{ref:diivine}, BTMQI \cite{ref:btmqi}, RcNet \cite{ref:rcnet}, HIGRADE \cite{ref:higrade}, and MSML \cite{ref:msml} are no-reference (NR) IQA models. We adapted all IQA models to videos by applying them framewise.

The median cross-validation accuracy achieved by each quality prediction model on the LIVE-TMHDR database is presented in Table \ref{tab:comparison}. From Table \ref{tab:comparison}, it may be seen that Cut-FUNQUE significantly outperformed, by over 15\%, nearly all of the compared quality models on the LIVE-TMHDR database in terms of quality prediction accuracy. The only existing quality model rivaling Cut-FUNQUE is the deep MSML model that computes a set of 9216 features from a pre-trained ResNet-50 model. By contrast, Cut-FUNQUE relies on an efficient Haar wavelet transform and computation sharing between features. 

To formally evaluate the efficiency of Cut-FUNQUE, we used the Performance Application Programming Interface (PAPI) \cite{ref:papi} to count the number of Floating Point Operations (FLOPs) required by each quality model evaluated in Table \ref{tab:comparison}. This measurement provides a theoretical estimate of each model's computational complexity. In addition, we also measured the running time of each quality model over 50 frames of a 1080p input video pair on a six-core Intel Core i7-8700 CPU, which has a clock frequency of 3.20GHz. Using this measurement, we report the average time taken per frame by each quality model as a dsecriptor of its practical computational complexity. These measurements, in units of Giga FLOPs and seconds, are also included in Table \ref{tab:comparison}. From this, it may be seen that Cut-FUNQUE achieves comparable accuracy as MSML, while being \(\sim 23 \times\) as efficient, in terms of FLOPs.

\begin{table*}[t]
    \centering
    \caption{Evaluation of Quality Prediction Models in Terms of Median Cross-Validation Accuracy and Computational Complexity}
    \label{tab:comparison}
    \begin{tabular}{|c|c|c|c|c|c|c|}
        \hline
         Model & Regressor & PCC & SROCC & RMSE & GFLOPs/Frame & Time (Sec/Frame) \\
         \hline
        Y-FUNQUE+ \cite{ref:funque_plus} & RandomForest & 0.4524 & 0.4343 & 9.4352 & 0.1042 & 2.414 \\
        BTMQI \cite{ref:btmqi} & GaussianSVR & 0.4705 & 0.4663 & 9.2238 & 0.1199 & 4.612 \\
        FSITM \cite{ref:fsitm} & LinearSVR & 0.4813 & 0.4626 & 8.9212 & 8.9487 & 4.930 \\
        NIQE \cite{ref:niqe} & GaussianSVR & 0.4805 & 0.4746 & 9.5563 & 2.3654 & 3.198 \\
        BRISQUE \cite{ref:brisque} & LinearSVR & 0.4811 & 0.4833 & 8.9869 & 0.2120 & 0.893 \\
        DIIVINE \cite{ref:diivine} & GaussianSVR & 0.4794 & 0.4925 & 9.2879 & 20.8771 & 155.373 \\
        TMQI \cite{ref:tmqi} & GaussianSVR & 0.5062 & 0.4956 & 8.6897 & 0.9061 & 2.374 \\
        FUNQUE \cite{ref:funque} & RandomForest & 0.5082 & 0.4949 & 8.8863 & 0.1716 & 2.471 \\
        TMVQI \cite{ref:tmvqi} & RandomForest & 0.5198 & 0.4969 & 8.8697 & 1.4164 & 3.002 \\
        FFTMI \cite{ref:fftmi} & GaussianSVR & 0.5298 & 0.5315 & 8.8559 & 27.5161 & 14.407 \\
        3C-FUNQUE+ \cite{ref:funque_plus} & RandomForest & 0.5817 & 0.5661 & 8.6568 & 0.3667 & 2.825\\
        RcNet \cite{ref:rcnet} & Random Forest & 0.5985 & 0.5824 & 8.2417 & 134.5597 & 1280.730 \\
        HIGRADE \cite{ref:higrade} & GaussianSVR & 0.6682 & 0.6698 & 8.2619 & 2.6533 & 3.709 \\
        \textbf{MSML} \cite{ref:msml} & Linear SVR & \textbf{0.7883} & 0.7740 & 6.8090 & 67.2578 & 412.438 \\
        \textbf{Cut-FUNQUE} & Random Forest & 0.7783 & \textbf{0.7781} & \textbf{6.4187} & 2.9257 & 6.343 \\
         \hline
    \end{tabular}
\end{table*}

To analyze the novel proposals in Cut-FUNQUE, we conducted an ablation study by comparing variants of Cut-FUNQUE that use different encoding functions (in Table \ref{tab:cut_funque_ablation_space}) and subsets of low-level cut weighted features (luminance-, spatial-, and temporal-complexity) (in Table \ref{tab:cut_funque_ablation_cuts}). From these results, it may be seen that replacing PUColor with PQ or PU21, or removing weighted-aggregated features decreases accuracy. These results validate the design choices made in Cut-FUNQUE.

\begin{table}[t]
    \centering
    \caption{Quality Prediction Accuracy of Cut-FUNQUE Variants Using Various Encoding Functions}\label{tab:cut_funque_ablation_space}
    \begin{tabular}{|c|c|c|c|}
    \hline
    Encoding Function & PCC & SROCC & RMSE \\
    \hline
    \textbf{PUColor} & \textbf{0.7783} & \textbf{0.7781} & \textbf{6.4187} \\
    PU21 & 0.7765 & 0.7686 & 6.4313 \\
    PQ & 0.7677 & 0.7619 & 6.4738 \\
    \hline
    \end{tabular}
\end{table}

\begin{table}[t]
    \centering
    \caption{Effect of Progressively Removing Weighted-Aggregated Features from Cut-FUNQUE}\label{tab:cut_funque_ablation_cuts}
    \begin{tabular}{|c|c|c|c|}
    \hline
    Model & PCC & SROCC & RMSE \\
    \hline
    \textbf{Cut-FUNQUE} & \textbf{0.7783} & \textbf{0.7781} & \textbf{6.4187} \\
    w/o T-Weighted Features & 0.7636 & 0.7540 & 6.4808 \\
    w/o L-weighted Features & 0.7513 & 0.7520 & 6.6442 \\
    w/o S-weighted Features & 0.7389 & 0.7337 & 6.9726 \\
    \hline
    \end{tabular}
\end{table}

Finally, we lend further weight to the comparison of quality prediction accuracy on LIVE-TMHDR by conducting a statistical significance test on the differences in accuracy. Specifically, we conducted one-sided Welch's t-tests to evaluate the statistical significance of the observed differences in prediction accuracies. A one-sided Welch's t-test was preferred over a traditional Student's t-test, since a Welch's t-test accounts for unequal population variances \cite{ref:welch}. Table \ref{tab:stat_comparison} presents the results of pairwise statistical significance comparisons. An entry of ``1'' (``0'') denotes that the quality model in the row achieved statistically significantly superior (inferior) accuracy compared to the quality model in the column. An entry of ``-'' denotes that the differences are not statistically significant. From this table, it may be seen that MSML and Cut-FUNQUE outperformed all other quality models, and that the difference between the two top performers was not statistically significant.

\begin{table*}[t]
    \centering
    \caption{Results of Welch's t-test Applied to Pairs of Objective Quality Models. \\ An entry of ``1'' (``0'') denotes that the model in the row achieves superior (inferior) accuracy to the model in the column, \\ while an entry of ``-'' denotes that the difference in accuracy is not statistically significant.}
    \label{tab:stat_comparison}
    \tiny
    \begin{tabular}{|c|ccccccccccccccc|}
    \hline
 & Y-FUNQUE+ & BTMQI & FSITM & NIQE & BRISQUE & DIIVINE & TMQI & FUNQUE & TMVQI & FFTMI & 3C-FUNQUE+ & RcNet & HIGRADE & MSML & Cut-FUNQUE\\
    \hline
    Y-FUNQUE+ & -  & -  & -  & -  & 0 & 0 & 0 & 0 & 0 & 0 & 0 & 0 & 0 & 0 & 0\\
    BTMQI & -  & -  & -  & -  & 0 & 0 & 0 & 0 & 0 & 0 & 0 & 0 & 0 & 0 & 0\\
    FSITM & -  & -  & -  & -  & -  & -  & 0 & 0 & 0 & 0 & 0 & 0 & 0 & 0 & 0\\
    NIQE & -  & -  & -  & -  & -  & -  & 0 & 0 & 0 & 0 & 0 & 0 & 0 & 0 & 0\\
    BRISQUE & 1 & 1 & -  & -  & -  & -  & -  & -  & -  & 0 & -  & 0 & 0 & 0 & 0\\
    DIIVINE & 1 & 1 & -  & -  & -  & -  & -  & -  & -  & 0 & -  & 0 & 0 & 0 & 0\\
    TMQI & 1 & 1 & 1 & 1 & -  & -  & -  & -  & -  & -  & -  & 0 & 0 & 0 & 0\\
    FUNQUE & 1 & 1 & 1 & 1 & -  & -  & -  & -  & -  & 0 & -  & 0 & 0 & 0 & 0\\
    TMVQI & 1 & 1 & 1 & 1 & -  & -  & -  & -  & -  & 0 & -  & 0 & 0 & 0 & 0\\
    FFTMI & 1 & 1 & 1 & 1 & 1 & 1 & -  & 1 & 1 & -  & -  & 0 & 0 & 0 & 0\\
    3C-FUNQUE+ & 1 & 1 & 1 & 1 & -  & -  & -  & -  & -  & -  & -  & -  & 0 & 0 & 0\\
    RcNet & 1 & 1 & 1 & 1 & 1 & 1 & 1 & 1 & 1 & 1 & -  & -  & 0 & 0 & 0\\
    HIGRADE & 1 & 1 & 1 & 1 & 1 & 1 & 1 & 1 & 1 & 1 & 1 & 1 & -  & 0 & 0\\
    MSML & 1 & 1 & 1 & 1 & 1 & 1 & 1 & 1 & 1 & 1 & 1 & 1 & 1 & -  & - \\
    Cut-FUNQUE & 1 & 1 & 1 & 1 & 1 & 1 & 1 & 1 & 1 & 1 & 1 & 1 & 1 & -  & - \\
    \hline
    \end{tabular}
\end{table*}

\section{Conclusion}
We have developed and validated a novel model called Cut-FUNQUE, which has been designed to predict the perceptual quality of tone-mapped and compressed HDR videos. An analysis of the quality prediction accuracy and computational complexity of Cut-FUNQUE demonstrates that it achieves comparable accuracy as the SOTA deep quality model MSML, while using a fraction (less than 5\%) of its computational cost.

Despite its benefits, Cut-FUNQUE could be improved further by applying more accurate global and local color and contrast quality features. This work also leaves room for studies of the trade-offs between reproducing the appearance of a source HDR video and the standalone quality of corresponding tone-mapped videos. Finally, the development of domain-specific deep neural networks that leverage knowledge of tone-mapping distortions may lead to further improvements in quality prediction accuracy. The primary challenge facing such approaches is the scarcity of publicly available HDR videos and even larger tone mapped video quality databases.

\label{sec:conclusion}

\bibliographystyle{IEEEtran}
\bibliography{refs}
\end{document}